\documentclass[11pt]{article}
\usepackage[skip=14pt plus2pt, indent=10pt]{parskip}
\usepackage{verbatim}
\usepackage{epsfig,graphics}
\usepackage {graphicx}
\usepackage {epsfig}
\usepackage {subfigure}
\usepackage {tabularx} 
\usepackage{rotate}	
\usepackage{slashed}
\usepackage{braket}

\usepackage{amsmath}
\usepackage{amsfonts}
\usepackage{amssymb}
\usepackage{graphicx}
\usepackage{xcolor}
\usepackage{cite}

\usepackage{fancyhdr}
\usepackage[hidelinks]{hyperref}

\usepackage{braket}
\usepackage{cancel}

\def\beq{\begin{equation}}
	\def\eeq{\end{equation}}
\def\beqa{\begin{eqnarray}}
	\def\eeqa{\end{eqnarray}}

\catcode`\@=11   
\newdimen\@rotdimen
\newbox\@rotbox  

\def\@vspec#1{\special{ps:#1}}
\def\@rotstart#1{\@vspec{gsave currentpoint currentpoint translate
		#1 neg exch neg exch translate}}
\def\@rotfinish{\@vspec{currentpoint grestore moveto}}
\def\@rotr#1{\@rotdimen=\ht#1\advance\@rotdimen by\dp#1%
	\hbox to\@rotdimen{\hskip\ht#1\vbox to\wd#1{\@rotstart{90 rotate}%
			\box#1\vss}\hss}\@rotfinish}
\def\@rotl#1{\@rotdimen=\ht#1\advance\@rotdimen by\dp#1%
	\hbox to\@rotdimen{\vbox to\wd#1{\vskip\wd#1\@rotstart{270 rotate}%
			\box#1\vss}\hss}\@rotfinish}%
\def\@rotu#1{\@rotdimen=\ht#1\advance\@rotdimen by\dp#1%
	\hbox to\wd#1{\hskip\wd#1\vbox to\@rotdimen{\vskip\@rotdimen
			\@rotstart{-1 dup scale}\box#1\vss}\hss}\@rotfinish}%
\def\@rotf#1{\hbox to\wd#1{\hskip\wd#1\@rotstart{-1 1 scale}%
		\box#1\hss}\@rotfinish}%
\def\rotate{\@ifnextchar[{\@rotate}{\@rotate[l]}}
\def\@rotate[#1]#2{\setbox\@rotbox=\hbox{#2}\@nameuse{@rot#1}\@rotbox}

\catcode`\@=12

\topmargin
-1.5cm
\textwidth
15.5cm
\textheight
23.5cm
\oddsidemargin
0.7cm
\evensidemargin
0.7cm

\begin{document}
	\makeatletter
	\@addtoreset{equation}{section}
	\makeatother
	\renewcommand{\theequation}{\thesection.\arabic{equation}}
	\pagestyle{empty}
	\leftline{IFT-UAM/CSIC-23-5}
	\vspace{-0.6cm}
	\rightline{\it To the memory of Graham Ross,}
	\rightline{\it dearest friend and colleague}
	\rightline{L.E.I.}
	\begin{center}
		
		\vspace{2cm}

		\def\Deq#1{\mbox{$D$=#1}}
		\def\Neq#1{\mbox{$N$=#1}}
		\def\preal{{\rm Re\,}}
		\def\pim{{\rm Im\,}}
		\def\ds{\displaystyle}
		\def\yzero{\smash{\hbox{$y\kern-4pt\raise1pt\hbox{${}^\circ$}$}}}
		\def\p{\partial}
		\def\a{\alpha}
		\def\b{\beta}
		\def\g{\gamma}
		\def\d{\delta}
		\def\beq{\begin{equation}}
			\def\eeq{\end{equation}}
		\def\beqa{\begin{eqnarray}}
			\def\eeqa{\end{eqnarray}}
		\def\Om{\Omega}
		\def\om{\omega}
		\def\th{\theta}
		\def\vt{\vartheta}
		\def\vphi{\varphi}
		\def\-{\hphantom{-}}
		\def\ov{\overline}
		\def\s2{\frac{1}{\sqrt2}}
		\def\wh{\widehat}
		\def\wt{\widetilde}
		\def\oh{\frac{1}{2}}
		\def\beq{\begin{equation}}
			\def\eeq{\end{equation}}
		\def\beqa{\begin{eqnarray}}
			\def\eeqa{\end{eqnarray}}
		\def\tr{{\rm tr \,}}
		\def\Tr{{\rm Tr \,}}
		\def\diag{{\rm diag \,}}
		\def\IF{\relax{\rm I\kern-.18em F}}
		\def\II{\relax{\rm I\kern-.18em I}}
		\def\IP{\relax{\rm I\kern-.18em P}}
		\def\IC{\relax\hbox{\kern.25em$\inbar\kern-.3em{\rm C}$}}
		\def\IR{\relax{\rm I\kern-.18em R}}
		\def\hm{\relax{n_H}}
		\def\vac{|0 \rangle}
		\def\vm{\relax{n_V}}
		\def\cc{{\cal C}}
		\def\ck{{\cal K}}
		\def\ci{{\cal I}}
		\def\cu{{\cal U}}   
		\def\cg{{\cal G}}
		\def\cn{{\cal N}}
		\def\cam{{\cal M}}
		\def\cp{{\cal P}}
		\def\ct{{\cal T}}
		\def\cv{{\cal V}}
		\def\cz{{\cal Z}}
		\def\ch{{\cal H}}
		\def\cf{{\cal F}}
		\def\tv{\tilde v}
		\def\Dsl{\,\raise.15ex\hbox{/}\mkern-13.5mu D} 
		\def\IZ{Z\kern-.4em  Z}
		\def\id{{\rm I}}
		
		
		\def\A{{\bf A}}   
		\def\B{{\bf B}}
		\def\ent{{\bf Z}}
		\def\C{{\bf C}}
		\def\ti{\times}
		\def\OR{\Omega {\cal R}}  
		\def\R{{\cal R}}
		\def\ca{{\cal A}}
		\def\lam{\lambda}
		\def\raw{\rightarrow}
		\def\Raw{\Rightarrow}
		\def\G{\Gamma}
		\def\ep{\epsilon}
		\def\arr{\arrowvert}
		\def\Arr{\Arrowvert}
		
		\LARGE{ Towers and Hierarchies in the Standard Model \\
			from Emergence in Quantum Gravity\\ [13mm]}

		\large{ A. Castellano$^1$, A. Herr\'aez$^2$ and L.E. Ib\'a\~nez$^1$
			\\[12mm]}
		\small{
			$^1$ {Departamento de F\'{\i}sica Te\'orica
				and Instituto de F\'{\i}sica Te\'orica UAM/CSIC,\\
				Universidad Aut\'onoma de Madrid,
				Cantoblanco, 28049 Madrid, Spain}  \\[5pt]
			$^2$  {Institut de Physique Th\'eorique, Universit\'e Paris Saclay, CEA, CNRS\\
				Orme des Merisiers, 91191 Gif-sur-Yvette CEDEX, France} 
			\\[12mm]
		}
		\small{\bf  Abstract} \\[8mm]
	\end{center}
	\begin{center}
		\begin{minipage}[h]{15.22cm}
			Based on Quantum Gravity arguments, it has been suggested that all kinetic terms of light particles below the UV cut-off could arise in the IR via quantum (loop) corrections. These loop corrections involve infinite towers of states becoming light (e.g. Kaluza-Klein or string towers). We study implications of this \emph{Emergence Proposal} for fundamental scales in the Standard Model (SM). In this scheme all Yukawa couplings are of order one in the UV and small Yukawas for lighter generations appear via large anomalous dimensions induced by the towers of states. Thus, the observed hierarchies of quark and lepton masses are a reflection of the structure of towers of states that lie below the Quantum Gravity scale, $\Lambda_{\text{QG}}$. Small Dirac neutrino masses consistent with experimental observation appear due to the existence of a tower of SM singlet states of mass $m_{0}\simeq Y_{\nu_3}M_p\simeq 7\times 10^5$ GeV, opening up a new extra dimension, while the UV cut-off occurs at $\Lambda_{\text{QG}}\lesssim 10^{14}$ GeV. Additional constraints relating the Electro-Weak (EW) and cosmological constant (c.c.) scales (denoted $M_{\text{EW}}$ and $V_0$) appear if the Swampland condition $m_{\nu_1}\lesssim V_0^{1/4}$ is imposed (with $\nu_1$ denoting the lightest neutrino), which itself arises upon applying the \emph{AdS non-SUSY Conjecture} or the \emph{AdS/dS Distance Conjecture} to the 3d vacua from circle compactifications of the SM. In particular, the EW scale and that of the extra dimension fulfill ${m_0 \, M_{\text{EW}}\lesssim 10^{2}\, V_0^{1/4}M_p}$, thus relating the EW hierarchy problem to that of the c.c. Hence, all fundamental scales may be written as powers of the c.c., i.e. $m_{\bullet}\sim V_0^\delta M_p^{1-4\delta}$. The scale of SUSY breaking is $m_{3/2}\lesssim 7\times 10^5$ GeV, which favours a \emph{Mini-Split} scenario that could be possibly tested at LHC and/or FCC.
		\end{minipage}
	\end{center}
	\newpage
	\setcounter{page}{1}
	\pagestyle{plain}
	\renewcommand{\thefootnote}{\arabic{footnote}}
	\setcounter{footnote}{0}
	

	
	\tableofcontents
	
	\section{Introduction}

	In spite of the amazing success of the Standard Model (SM) of Particle Physics in describing all available experimental data, our total lack of understanding of the mass scales in which it is based upon is quite embarrassing. In the present paper we try to improve this understanding, so let us begin by recalling the observed structure of fundamental scales in physics:
	
	\begin{itemize}
		
		\item The less understood scale from a fundamental point of view is that of the cosmological constant (c.c.), whose existence provides the simplest explanation for the accelerated expansion of the universe. Its value is $V_0\simeq (2.4\ \text{meV})^4$, around 120 orders of magnitude smaller than the largest fundamental scale $\sim M_p^4$, with $M_p$  the (reduced) Planck mass $M_p\simeq 2.4 \times 10^{18}$ GeV.
		
		\item  The Electro-Weak (EW) scale, $M_{\text{EW}}\simeq 10^2\ \text{GeV}\simeq 10^{-16}M_p$. Here the issue is that it is fixed by the vacuum expectation value (VEV) of a Higgs scalar $\braket{H_0} \simeq 175$ GeV, whose mass is unstable against quadratic loop corrections. Therefore we do not understand how this huge hierarchy is maintained. An attractive explanation to ensure this stability is the possible presence of low-energy supersymmetry (SUSY), slightly above the EW scale. However LHC data has shown no trace of supersymmetric particles so far and a certain degree of fine-tuning is already required. Furthermore, even if SUSY was found, solving this stability problem, an understanding of the hierarchy itself would still be required. An interesting relationship which approximately matches with the observed values is the following
	    \beq
		\tilde M\, \sim\, \sqrt {V_0^{1/4}M_p}\, \simeq\, 10^3\,  \text{GeV}\, ,
		\label{eq:EW&cc}
		\eeq
		in which  a scale $\tilde M$ slightly above the EW scale appears as a geometric mean of the c.c. and Planck scales. However, no \emph{compelling} theoretical derivation of such an expression is currently available.
		
		\item Neutrino masses are much smaller than that of any other charged fermion in the SM. Oscillation experiments have told us that two neutrinos have masses in a range about $m_{\nu_{2,3}}\sim 10^{-1}-10^{-2}$ eV, whereas the lightest neutrino may be arbitrarily light. It is not yet known if the observed neutrinos have Majorana or Dirac masses and whether these present a normal or an inverted hierarchy (although the former is slightly favored). As in the expression above, there is the remarkable numerical fact for the overall scale of neutrinos 
		\beq
		m_\nu\, \sim\, V_0^{1/4}\, .
		\label{eq:neutrinomass&cc}
		\eeq
		These small values for neutrino masses can be straightforwardly explained in terms of the see-saw mechanism
		\cite{Minkowski:1977sc,Gell-Mann:1979vob,Yanagida:1980xy},
		in which they get Majorana masses of order $m_\nu\sim |\braket{H_0}|^2/M_{\text{L}}$, with $M_{\text{L}}$ some high scale at which lepton number is violated, but the mechanism itself does not explain the interesting coincidence in eq. \eqref{eq:neutrinomass&cc}.
		
		\item The particular value of the Higgs mass has also raised some new questions. With ${m_{\text{H}}\simeq 126 \ \text{GeV}}$, a renormalisation group computation for the (non-SUSY) SM tells us that the Higgs potential becomes unstable (or metastable) at a mass scale of the order of $10^9-10^{12}$ GeV\cite{Elias-Miro:2011sqh,Degrassi:2012ry}. One way to avoid this instability is the presence of SUSY at a scale $\lesssim 10^{12}$ GeV. In fact, the experimental value of the Higgs mass is consistent with the predictions of the Minimal Supersymmetric Standard Model (MSSM) which yield $m_{\text{H}}\lesssim 140$ GeV, with the upper bound being only possible for a very massive SUSY spectrum. Indeed, a value of $m_{\text{H}}\simeq 126$ GeV requires heavy squarks with masses typically $\gtrsim 10^4$ GeV. Thus, this may point to the possible presence of SUSY in a range $M_{\text{SUSY}}\sim 10^4\sim 10^{12}$ GeV.
		
	\end{itemize}
	
	Additionally, the flavour structure is not yet  understood, although some general patterns are apparent: The spectrum of quarks and charged leptons is hierarchical, with $m_{\text{gen}\, ,1}\ll m_{\text{gen}\, ,2}\ll m_{\text{gen}\, ,3}$ for the three generations and an approximately diagonal CKM matrix. These general patterns may be obtained in a variety of models, see e.g. \cite{Feruglio:2015jfa} for a review.
	
	In most traditional approaches to address the above structure of fundamental scales, possible Quantum Gravity (QG) effects are usually neglected. In principle, this looks like a reasonable attitude from the Wilsonian point of view, since the Planck scale $M_p\simeq 10^{18}$ GeV is well above any relevant scale involved in SM physics, and therefore one expects any local operator involving quantum gravity to be suppressed by powers of the Planck mass. However, it is becoming increasingly clear that in addressing aspects like e.g. the EW hierarchy problem or the c.c. issue this approach may be unjustified. The tacit assumption being that any (anomaly-free) low-energy quantum field theory one can write down may be consistently coupled to QG. However, this assumption is incorrect and it has been argued that \emph{most} of the effective field theories (EFTs) one can think of cannot be consistently coupled to gravity in the UV \cite{Vafa:2005ui}. String theory, our firmest candidate for such a consistent QG theory, provides us with many such examples, like e.g. the fact that in $d=10$ there are only a few consistent theories that correspond to the known 10d SUSY-strings (and a few non-SUSY ones). The identification of the theories that are consistent versus those which are not is the aim of the Swampland Program (see \cite{Palti:2019pca,vanBeest:2021lhn,Grana:2021zvf} for reviews). 
	
	In this paper we try to study how some of the recent developments in the Swampland Program may shed new light on our understanding of the structure of fundamental scales in physics. One of the ideas put forward within this context is that of \emph{Emergence}  \cite{Palti:2019pca,Harlow:2015lma,Grimm:2018ohb,Corvilain:2018lgw,Heidenreich:2017sim,Heidenreich:2018kpg,Blumenhagen:2019qcg, Blumenhagen:2019vgj, EnriquezRojo:2020hzi, Blumenhagen:2020dea, Castellano:2022bvr,Marchesano:2022axe}. In a nutshell, it amounts to the statement that the  kinetic terms of all massless particles are negligible at the UV scale of the theory. They should then appear as an IR effect due to loop corrections involving the sum over infinite towers of states becoming light (in Planck units) with a characteristic mass scale $m_0$. These towers correspond to e.g. Kaluza-Klein (KK) towers or in some cases to string oscillators.\footnote{See \cite{Castellano:2022bvr} for a recent systematic study of this Emergence Proposal in quantum gravity and some string theory tests.} In this article we emphasise that, in the context of Emergence, hierarchies of Yukawa couplings may appear easily. Indeed, one interesting point within this paradigm is that if e.g. a fermion couples to one such tower, its wave-function $\mathcal{Z}$ gets enhanced by factors of the form $\mathcal{Z}\sim (\Lambda_{\text{QG}}/m_0)^\gamma$, with $\gamma$ some constant of order one and $\Lambda_{\text{QG}}$ denotes the UV cut-off \cite{Castellano:2022bvr}. This is interesting because, once one goes to canonical kinetic terms for the light fermions, Yukawa couplings $Y_0^{ijk}$ get suppression factors $Y_0^{ijk}\to \left(\mathcal{Z}_i \mathcal{Z}_j \mathcal{Z}_k\right)^{-1/2}Y_0^{ijk}$, which may then generate hierarchies of fermion masses, even if we started with all Yukawa couplings $Y_0^{ijk}$ at the UV being of order one. We discuss how this structure may describe the observed mass hierarchies of charged leptons and quarks in the SM, assuming the latter couple to appropriate towers of states. Such a situation, in which SM hierarchies are generated via hierarchical kinetic terms, has a long history (see e.g. \cite{Georgi,Ibanezcoset,Nelson,Poland,Dudas}). The structure of SM mass  hierarchies we obtain is somewhat similar to the one in the `Superconformal Flavour' models studied in \cite{Nelson,Poland}, where the existence of strongly interacting superconformal sectors enhance the wave-function renormalisation of fermions by factors of the form $\mathcal{Z}\sim (\Lambda_{\text{UV}}/\Lambda_{\text{c}})^\gamma$, with $\gamma$ being some anomalous dimension of order one and $\Lambda_{\text{c}}$ the scale below which the strongly interacting sector ceases to be conformal \cite{Nelson,Poland}. In spite of this formal similarity, the physics underlying the two schemes is quite different, since in our case the hierarchies originate from the existence of towers of states --- of quantum gravitational nature --- coupling to massless matter fields, which is a characteristic feature of string theory compactifications.
	
	The generation of hierarchies via towers of light states gets particularly interesting in the neutrino sector. In particular, if right-handed neutrinos 
	exist and have very large wave-function renormalisation due to their couplings to a light tower of states of characteristic scale $m_0$, the neutrinos may get hierarchically small Yukawa couplings $Y_{\nu}\propto m_0/M_p$. We find that Dirac masses consistent with experimental neutrino data are obtained if $m_0\simeq 7\times 10^5$ GeV, signaling the opening of an extra dimension above that scale. The existence of such a tower of states implies a lowering of the fundamental gravitational scale from the naive 4d Planck mass, $M_p$, down to $\Lambda_{\text{QG}}\simeq 10^{14}$ GeV.\footnote{Here $\Lambda_{\text{QG}}$ is the `species scale' defined as the effective scale at which quantum gravity effects become dominant \cite{Dvali:2009ks,Dvali:2007hz,Dvali:2010vm}, and it captures the fundamental QG scale of the UV complete theory.} Thus, in this scheme neutrinos are naturally of Dirac type and their observed masses are very small because the right-handed components receive large wave-function renormalisation effects through their coupling to such an infinite tower of states.	
	
	With the presence of a tower at $m_0 \simeq 7\times 10^{5}$ GeV one can describe the smallness of neutrino masses but this by itself does not give us a reason for why such a low scale exists in the first place. However, more constraints about the fundamental scales of physics are obtained if one assumes the validity of certain Swampland criteria, in particular the \emph{AdS non-SUSY conjecture} \cite{Ooguri:2016pdq} and/or the {\it AdS Distance Conjecture } (ADC) of ref.\cite{Lust:2019zwm}. The first one, which has been tested in many string theory examples, states that stable non-SUSY AdS vacua cannot exist in QG. It is known that if we compactify the SM on a circle, AdS minima tend to appear in 3d originated by Casimir forces balanced against the  term coming from the pure 4d vacuum energy, $V_0$ \cite{Arkani-Hamed:2007ryu}. Avoiding such problematic three-dimensional AdS vacua leads to an upper bound on the lightest neutrino mass in terms of the 4d c.c. \cite{Ibanez:2017kvh,Ibanez:2017oqr,Hamada:2017yji} (see also \cite{Gonzalo:2018tpb,Gonzalo:2018dxi,Gonzalo:2021zsp}):
	\beq
	m_{\nu_1}\, \lesssim\, V_0^{1/4}\, .
	\label{eq:neutrinoineq}
	\eeq 
	Essentially the same condition is obtained in terms of the ADC, which in this case states that within said 3d SM there cannot be transitions from positive to negative energy minima \cite{Gonzalo:2021zsp}. In any event, the above inequality is an interesting bound which would provide for an understanding of the experimental coincidence between the scale of neutrino masses and that of the c.c., as pointed out in eq. \eqref{eq:neutrinomass&cc}. As we argue in the present work, this leads to a bound of the form $m_0\, M_{\text{EW}} \lesssim 10^2\times V_0^{1/4}M_p$, which in order not to be violated requires again that $m_0\lesssim  10^6$ GeV. Thus, we now have an understanding of why neutrinos are so light: We conclude that higher values for neutrino masses would violate the aforementioned Swampland conjectures. One also finds separate upper bounds for both the EW scale and the tower scale $m_0$
	\beq
	M_{\text{EW}}\, \lesssim\, \epsilon^{1/2}\, V_0^{1/8}\, M_p^{1/2}\, \sim\, 10^2\, \text{GeV}\, , \qquad m_0\, \lesssim\, 10 \times \epsilon^{-1/2}\, 
	V_0^{1/8}\, M_p^{1/2}\, \sim\, 7\times 10^5\, \text{GeV}\, ,
	\eeq
	where $\epsilon\sim 10^{-2}$ is related to the mass hierarchy of charged leptons (see below). The first of these expressions provides for a solution to the EW hierarchy assuming the above Swampland conditions hold. Notice that we now obtain an expression like the one in eq. \eqref{eq:EW&cc} upon defining $\tilde M$ as the geometric average between the EW and the tower scales, i.e. ${\tilde M}=\sqrt{M_{\text{EW}}\, m_0}\,$, which does verify ${\tilde M} \sim \sqrt{V_0^{1/4}M_p}$. 
	
	Thus we find that a combination of Emergence and the AdS non-SUSY (and/or  the ADC) conjectures would provide for an understanding of the EW and neutrino mass scales in terms of the c.c., $V_0$. Emergence would also provide for a mechanism for the generation of quark and lepton mass hierarchies. In all this argumentation supersymmetry does in fact not play any crucial role, although one may still ask whether SUSY would be compatible with this structure. One can give an upper bound on the gravitino mass by noting that $m_{3/2}\simeq |W|/M_p^2\lesssim \Lambda_{\text{QG}}^3/M_p^2\simeq  7\times 10^5$ GeV, where we have assumed that the maximum VEV for the superpotential is bounded by $\Lambda_{\text{QG}}^3$, with $\Lambda_{\text{QG}}\simeq 10^{14}$ GeV being the UV cut-off. Thus one expects typical SUSY masses below $M_{\text{SS}}\lesssim 7\times 10^5$ GeV. Interestingly it is known that a typical SUSY spectrum with squarks in the range $10^4-10^6$ GeV leads to a Higgs mass around $m_{\text{H}}\sim 126$ GeV and is thus consistent with LHC data. More specifically such values for the SUSY spectrum are characteristic of the \emph{Mini-Split} scenario in \cite{Arvanitaki:2012ps,Arkani-Hamed:2012fhg,Hall:2012zp}, so that much of the phenomenology of these models would also apply to our setting.
	
	One remarkable observation to be made is that all fundamental scales in the theory, including the UV scale, $\Lambda_{\text{QG}}$, the EW scale, $M_{\text{EW}}$, the tower scale, $m_0$, the neutrino masses and the gravitino mass --- in the case of a supergravity EFT --- can be written in terms of powers of the c.c., namely $m_{\bullet}\sim V_0^{1/n}M_p^{1-4/n}$, with $n=24,8,4$. Thus, all scales go to zero as $V_0\rightarrow 0$. This is in agreement with the so-called \emph{AdS Distance Conjecture} \cite{Lust:2019zwm} which states that in the limit $|V_0|\rightarrow 0$ an infinite tower of states must appear, which in the present case would be fulfilled by the tower of scale $m_0$. It is thus natural to consider an {\it Upside Down Universe} \cite{PASCOS2022} in which the fundamental scale is the cosmological constant. In this context we briefly discuss the possibility of the c.c. being an external boundary condition in the theory, as suggested in \cite{Banks}.
	
	The structure of the rest of the paper is as follows. In the next section we introduce the Emergence Proposal in Quantum Gravity and how hierarchies of Yukawa couplings naturally arise in that context. Additionally, we briefly review some ideas from the Swampland Program which are used later. In section \ref{s:EmergenceSM} we discuss how the mass hierarchies of the Standard Model fermions, including neutrinos, appear in this scheme. In section \ref{s:Hierarchy&cc} we review how the Swampland bound \eqref{eq:neutrinoineq} leads to an upper bound on the EW scale and how all fundamental scales may be expressed in terms of powers of the c.c., $V_0$. We also discuss the (minimal) supersymmetric SM case and some phenomenological implications. Finally, in section \ref{s:ccUniverse} we present a short digression about the new perspectives that may arise from our analysis, emphasizing the idea that all low energy scales in the Standard Model seem to depend on the cosmological constant. We conclude with some final comments and open questions in section \ref{s:Outlook}.
	
	\section{Emergence and the Swampland}
	\label{s:Emergence&Swampland}
	
	In this section we introduce some ideas about the Swampland Program which will be helpful in the sections below. A first point to remark is that \emph{in the presence of Quantum Gravity}, i.e. for a finite value of $M_p$, some unfamiliar phenomena may occur when coupling an EFT to the gravitational field. A simple example is a $U(1)$ theory coupled to charged matter with gauge coupling $g$. In field theory not coupled to gravity one can take the limit $g\to 0$ and nothing special happens. On the contrary, one would say that perturbation theory gets better and better. However, in the presence of gravity we have learned in the last decade that as $g\to 0$ there is a cut-off, $m_0$, which behaves as
	\beq
	g\to 0\, \Rightarrow\,  m_0\, \lesssim\, g\, M_p\, .
	\label{eq:mWGC}
	\eeq
	This physics cut-off, $m_0$, turns out to correspond to an infinite tower of states, in particular KK or string excitations, and $m_0$ is typically the mass of the lightest state in the tower. Note that this effect disappears in the absence of quantum gravity, i.e. as $M_p\rightarrow \infty$. This property goes under the name of \emph{magnetic Weak Gravity Conjecture} (WGC) \cite{Arkani-Hamed:2006emk} (see also  \cite{Palti:2019pca,vanBeest:2021lhn,Grana:2021zvf, Harlow:2022gzl}). Heuristically, if $g$ becomes too small there is the risk of the gauge interaction becoming weaker than gravity, and QG tries to avoid this by taking the theory towards a higher dimension or popping in critical string excitations. The presence of \emph{towers} of states is ubiquitous in QG and string theory. In this context, it has also been shown in large classes of string vacua that, considering a geodesic distance  $\Delta_{\phi}$ in moduli space larger than the Planck scale, a tower of states with exponentially dumped masses appears, namely $m_{\text{tower}}\sim m_0\, e^{-\lambda \Delta_{\phi}}$, with $\lambda$ some $\mathcal{O}(1)$ constant. This has been termed as the \emph{Distance Conjecture} \cite{Ooguri:2006in,Ooguri:2018wrx,Grimm:2018ohb,Font:2019cxq}.
	
	While these phenomena are well established and have been checked in large classes of string vacua, it has also been conjectured that, if there is a family of scalar potentials with minima $|V_0|\to 0$, a tower of states with characteristic mass $m_{\text{tower}}$ should appear with
	\beq
	|V_0|\to  0\, \Rightarrow\,  m_{\text{tower}}\, \lesssim\, |V_0|^{\alpha}\, M_p^{1-4\alpha} \, ,
	\label{eq:ADC}
	\eeq
	where $\alpha$ is an $\mathcal{O}(1)$ positive constant whose range of allowed values may depend on the sign of $V_0$\cite{Lust:2019zwm}.\footnote{This conjecture is deeply related with the Distance Conjecture, as it arises from a generalization of the notion of infinite distances from moduli spaces to general field configurations (see \cite{Lust:2019zwm} for details).} This has been tested for AdS in plenty of string vacua  but only conjectured for dS vacua. The latter are notoriously difficult to construct in string theory\cite{Obied:2018sgi,Ooguri:2018wrx,Garg:2018zdg}, and hence there is less explicit evidence for the above conjecture in positively curved backgrounds. If applied to our observed universe, identifying $V_0$ with the observed cosmological constant, and given its tiny value, it means that there should be an infinite tower of \emph{light} (compared to the fundamental scale) particles in the theory which describes our universe. However, the scale of this tower depends sensitively on the (unknown) value of the $\alpha$ parameter. We will argue below that matching of neutrino mass data suggests ${ m_0\simeq 7\times 10^5\ \text{GeV}}$ and $\alpha=1/8$ in the context of Emergence.
	
	These phenomena  are unfamiliar from the pure EFT point of view. They, as well as other generic properties of QG, are  the subject of systematic scrutiny at the moment. One of the main goals being to identify whether there is some general principle underlying the different properties, some unifying scheme. One proposal along this direction is the \emph{Emergence Proposal}, which we describe next.
	
	\subsubsection*{The Emergence proposal}
	\label{emergencia1}
	
	The Emergence Proposal\cite{Palti:2019pca,Harlow:2015lma,Grimm:2018ohb,Heidenreich:2017sim,Heidenreich:2018kpg,Castellano:2022bvr}, may be formulated --- in its strongest version --- as follows:

	\emph{In a theory of Quantum Gravity all light particles in a perturbative regime  have no kinetic terms in the UV. The required kinetic terms appear as an IR effect after integrating out infinite towers of massive states below the QG cut-off.}

\begin{figure}[tb]
		\begin{center}
			\includegraphics[scale=0.45]{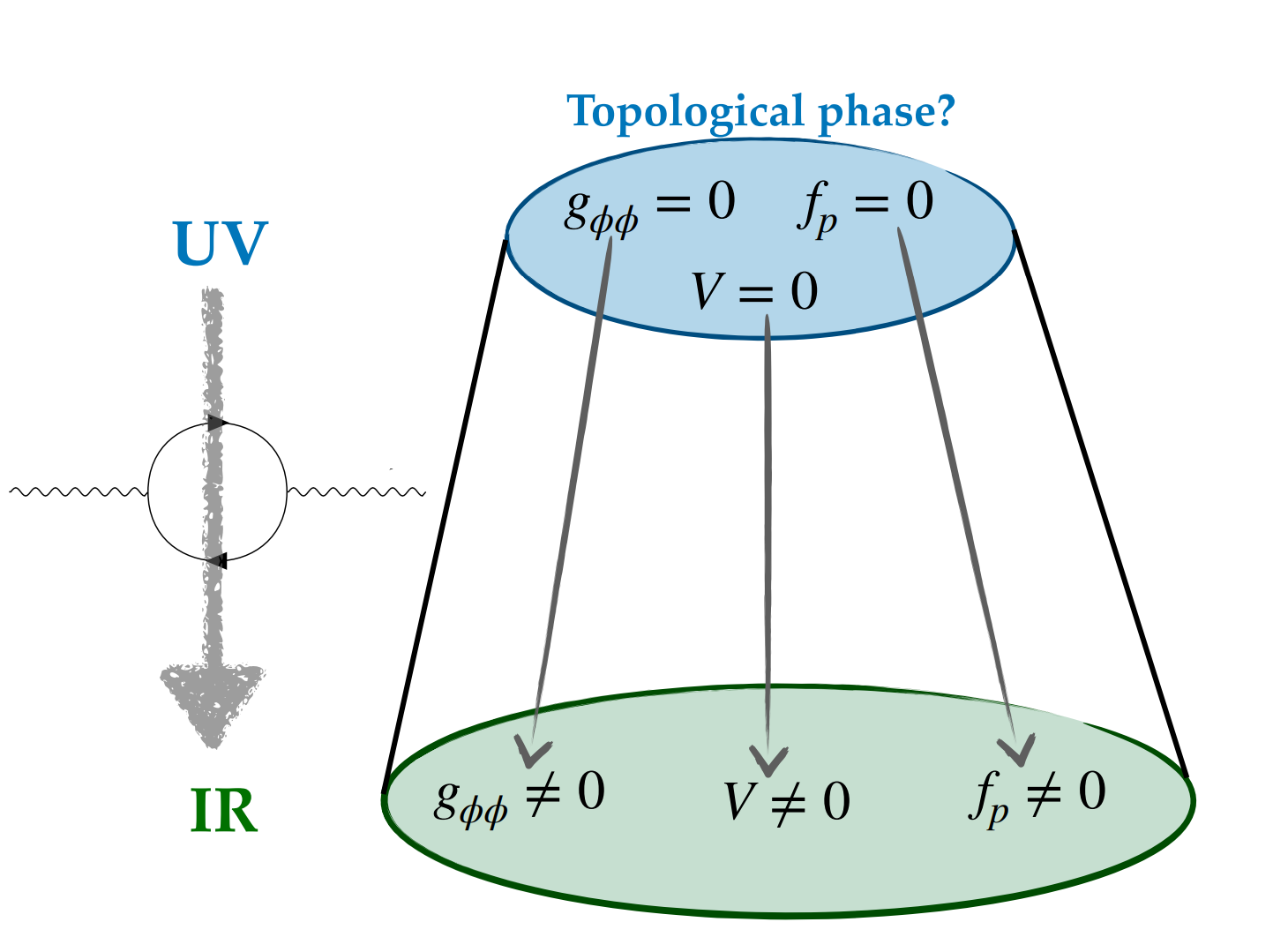}
			\caption{Scheme of  Emergence in QG. Massless fields do not have any kinetic term in an underlying strongly coupled theory. The metrics are generated by loops  involving the contributions of towers of states.}
			\label{strongemergence}
		\end{center}
	\end{figure}

	Note that the absence of kinetic terms implies that the theory in the UV must be strongly coupled, and the observed weakly coupled SM should appear as an IR effect. It has been argued that the condition of vanishing kinetic terms in the UV could perhaps suggest the existence of an underlying topological fundamental theory, in which particles do not propagate (see fig. \ref{strongemergence}). There would be couplings among non-propagating fields  but no geometric objects, i.e., no metrics for the kinetic terms, see e.g.\cite{Harlow:2015lma,Agrawal:2020xek}. An attractive feature of this proposal is that it has been argued to provide for a qualitative understanding of both the case of the magnetic WGC and the SDC mentioned above. 
	
	Let us first discuss how  this arises for the WGC. Consider a four-dimensional $U(1)$ gauge theory with no kinetic term for its massless fields to begin with, but with (minimal) couplings to a tower of states with increasing masses and (quantized) charges given by
	\beq
	m_n^2\, =\, n^2 m_0^2\, , \qquad q_n\, =\, n\, ,
	\label{torres}
	\eeq
	with $n \in \mathbb{Z}$. This correspond to a standard single KK tower characterized by the scale $m_0$. An important definition  is that of the \emph{species scale}, denoted here as $\Lambda_{\text{QG}}$, which gives the UV cut-off scale at which QG effects become important. It is given by \cite{Dvali:2009ks,Dvali:2007hz,Dvali:2010vm}
	\beq
	\Lambda_{\text{QG}}^2\, \lesssim\, \frac {M_p^2}{N}\, ,
	\eeq
	where $M_p$ is the Planck scale and $N$ is the number of degrees of freedom below $\Lambda_{\text{QG}}$ itself. Note that the highest state of the tower that we can consider is the one whose mass sits precisely around $\Lambda_{\text{QG}}$, i.e. $\Lambda_{\text{QG}}\simeq N\, m_0$. Consider now the one-loop contribution due to this tower of charged particles to the $U(1)$ wave-function renormalisation. It is computed as follows 
	\beq\label{eq:U(1)gaugeEmergence}
	\frac {1}{g^2}\, \simeq\, \sum_n^N n^2\, \text{log} \left(\frac {\Lambda_{\text{QG}}^2}{m_0^2\, n^2}\right)\, \sim\,  N^3\, \lesssim\, \frac {M_p^2}{m_0^2}\, ,
	\eeq
	where we have approximated the summation with an integral (which is justified whenever the number of states is sufficiently large, as in the relevant cases here) and we used the relation $\Lambda_{\text{QG}}\simeq N\, m_0$ as well as $\Lambda_{\text{QG}}^2\lesssim M_p^2/N$. We then obtain that the scale of the tower is bounded as $m_0^2\lesssim g^2M_p^2$, which is the magnetic WGC bound in eq. \eqref{eq:mWGC}. Thus the one-loop contribution from the tower gives rise to a kinetic term for the gauge vector bosons but also in a way which is automatically consistent with the WGC. Similarly, the exponentially suppressed behaviour of the masses of the towers of states with the distance in moduli space can be derived, as required by the SDC.

\begin{figure}[t]
		\begin{center}
			\subfigure[]{
				\includegraphics[height=3cm]{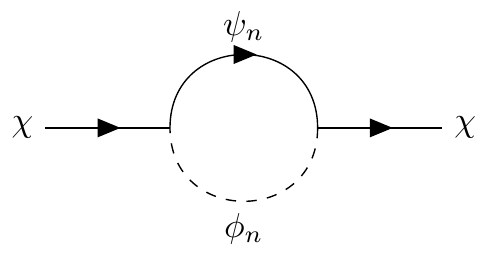}
			}\qquad \qquad
			\subfigure[]{
				\includegraphics[height=3cm]{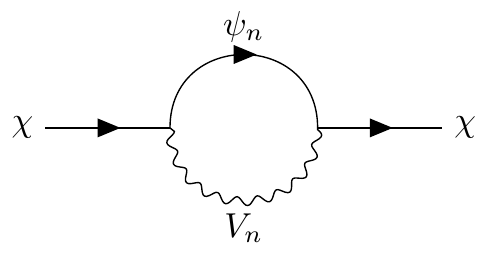}
			}
			\caption{One-loop diagrams contributing to the kinetic terms of fermions.}
			\label{kineticfermionsbas}
		\end{center}
	\end{figure}

	Let us consider now the one-loop generation of kinetic terms for light fermions.\footnote{Additional constraints due to the presence of fermionic particles in consistent low energy EFTs were already explored in \cite{Palti:2020tsy}. In that work, non-trivial consequences for the Yukawa couplings and the scale of supersymmetry breaking were proposed.} This is different from the case of gauge bosons for a couple reasons. In particular, the loops in question must contain one fermion and one boson in the internal propagators, interacting with the massless fermion through couplings as the ones displayed in fig. \ref{kineticfermionsbas}. That means that we need, in general, towers of massive fermions as well as towers of massive bosons, with couplings of the form $Y_n\ \overline{\phi^{(n)}} \left(\psi^{(n)}\chi \right)$  and/or ${\tilde Y}_n \left(\psi^{(n)} \sigma^{\mu} \overline \chi \right) V_{\mu}^{(n)}$, where $\phi^{(n)}$ and $V_{\mu}^{(n)}$ denote the towers of scalars and vector bosons, respectively. In general, the towers that enter each loop may be independent and, unlike in the case of gauge interactions, the Yukawa couplings $Y_n,{\tilde Y}_n$ are not a priori restricted by gauge invariance. As a proof of concept, we consider the simplest case, in which the towers that run in the loop have both the same mass structure; and the Yukawa couplings of the massless fermion, $\chi$, to the massive modes scale as $Y_n,\ {\tilde Y}_n \sim n$, similarly to the case of the gauge bosons considered above. This is the typical behavior when e.g. the massless fermion is a gaugino in an $\mathcal{N}=1$ theory. One may consider also the case in which the boson and fermion towers in the loop have different mass scales (see \cite{Castellano:2022bvr}), but the qualitative structure remains the same. Therefore, we take the masses for both fermions and bosons in the towers to be equal, namely $m_n^{\text{b}}=m_n^{\text{f}}=m_n$, and
	\beq
	m_n\, =\, n\, m_0\, , \qquad  \Lambda_{\text{QG}}\, \simeq\, N_0\, m_0\, , \qquad Y_n\, =\, n \, ,
	\eeq
	where $N_0$ is the total number of states in the towers. Then one gets for the quantum-corrected metric of the massless fermion after summing over the tower of states the following
	\beq
	g_{\chi \chi}\, \sim\, \sum_n^{N_0}  n^2\, \sim\, N_0^3\, \simeq\, \left(\frac {M_p}{m_0}\right)^2\, ,
	\eeq
	where we have used the species scale definition $\Lambda_{\text{QG}}^2\simeq M_p^2/N_0$. It is interesting for later use to consider additional independent --- or \emph{additive}, as explained in \cite{Castellano:2022bvr} --- towers\footnote{From a string theory perspective, one may think of these subleading towers as living on a localised defect, such as a brane, whilst the dominant one may be associated to a transverse direction in the bulk.} labeled by an index $i \in \mathbb{N}$ with
	\beq
	m_n^{(i)}\, =\, n_i\, m_0^{(i)}\, , \qquad Y_n^{(i)}\, =\, n_i\, ,
	\label{eq:mass&Yuksmalltowers}
	\eeq
	assuming also $m_0^{(i)}\gg m_0$, and hence with much fewer total number of states, $N_i\ll N_0$. In such a case the species scale is approximately the same,
	since $\Lambda_{\text{QG}}^2\sim M_p^2/(N_0+\sum_iN_i)\sim M_p^2/N_0$. We have 
	\beq
	\Lambda_{\text{QG}}\, \simeq\, N_0\, m_0\, \simeq\, N_i\, m_0^{(i)}\, , 
	\eeq
	and then for the metric, $g_{i \bar i}$, of the massless chiral fermions to which the heavier towers couple we find
	\beq\label{eq:wavefunctionheavytowers}
	g_{i \bar i}\, \sim\,  \sum_{n_i}^{N_i} n_i^2\, \sim\, N_i^3\, \sim\, \left(\frac {\Lambda_{\text{QG}}}{m_0^{(i)}}\right)^3\, .
	\eeq
	Thus, one observes that the wave-function renormalisation due to both types of towers, heavy and light, may become quite large if the tower scales, $m_0$ and $m_0^i$, also become small. We will take profit of this latter fact to generate hierarchical Yukawa couplings in the SM below.
	
	Notice that in theories with extra dimensions and string compactifications the masses of the particles in the loops depend on certain moduli, $m_{\text{tower}}=m_{\text{tower}}(\phi)$, so what one actually obtains is moduli dependent kinetic terms $g_{i \bar j}(\phi)$. Thus e.g. in a circle compactification of a 5-th dimension of radius $R_5$ one would have $m_0\simeq 1/R_5$ and a quantum-induced kinetic term $g_{\chi \chi}\sim R_5^2$.

	\subsubsection*{Hierarchies and  Yukawa couplings}
	
	Let us consider now some renormalisable Yukawa coupling and see what is the effect of this emergent behaviour.  To simplify notation let us consider a $\mathcal{N}=1$ case with a superpotential $W(\Phi)={\cal S}_{ijk}\Phi^i\Phi^j\Phi^k$, although the results will still apply in a non-SUSY setting. The kinetic terms are given by metrics $g_{i \bar i}$ which we take here to be diagonal so as to simplify the discussion. Then, after going to canonical kinetic terms for the light fields one gets physical Yukawa couplings of the form
	\beq
	Y_{ijk}\ =\ {\cal S}_{ijk} (g_{i \bar i}g_{j \bar j}g_{k \bar k})^{-1/2} \ .
	\eeq
	The idea in Emergence is that all entries in ${\cal S}_{ijk}$ are $\mathcal{O}(1)$ numbers and any possible \emph{hierarchically} small Yukawa coupling should come from the dependence on the metrics. Let us consider the case of the towers defined in	 eq. \eqref{eq:mass&Yuksmalltowers}. One would get for the Yukawa couplings a structure
	\beq
	Y_{ijk}\ \simeq \ {\cal S}_{ijk}  \left[  \left( \frac{m_0^{(i)}}{\Lambda_{\text{QG}}}\right) \left( \frac{m_0^{(j)}}{\Lambda_{\text{QG}}}\right) \left( \frac{m_0^{(k)}} {\Lambda_{\text{QG}}}\right) \right]^{3/2}\, ,
	\eeq
	with $\Lambda_{\text{QG}}$ the relevant UV (species) scale. It is clear that the size of the Yukawa couplings will get suppression factors whenever any light field $\Phi^i$ couples to a tower of states of characteristic mass scale $m_0^{(i)}$. We will make use of this feature to obtain hierarchies of SM Yukawa couplings in the next section.
	
	It is worth remarking at this point the similarity between the above generation of hierarchical Yukawas and that of the `Superconformal Flavour Models' pioneered in \cite{Nelson,Poland}.\footnote{Perhaps this similarity would  not be a coincidence if an holographic dual of the SCFT existed, which would live in one higher dimension, similar to our setting.} In those models the matter fields interact with a conformal sector and different generations present different anomalous dimensions.\footnote{Our convention for the anomalous dimension is $\gamma \equiv - \mu \frac{\partial \log \mathcal{Z}}{\partial \mu}$, following that of the references\cite{Nelson,Poland}.} Thus, the wave-function factors $\mathcal{Z}_i$ for each field $\Phi^i$ obey 
	\beq
	\mathcal{Z}_i^{-1/2}\,  =\,  \text{exp} \left(-\frac {1}{2} \int_{\text{log}\, \Lambda_{\text{c}}^{(i)}}^{\text{log}\, \Lambda_{\text{UV}}} \gamma_i\, \text{d}(\text{log}\,\mu)\right)\, \simeq\,  \left(\frac {\Lambda_{\text{c}}^{(i)}}{\Lambda_{\text{UV}}}\right)^{\frac {1}{2}\gamma_i}\, ,
	\eeq
	where $\Lambda_{\text{c}}^{(i)}$ is the scale at which the strongly coupled theory connected with $i$-th particle ceases to be conformal and $\gamma_i$ is the anomalous dimension which we take to be approximately constant and naturally of order one. In the simplest models one has a single conformal sector and there is a single conformal scale $\Lambda_{\text{c}}^{(i)}=\Lambda_{\text{c}}$. Interestingly, we see that in our case we get the same suppression factor which would correspond to an anomalous dimension $\gamma_i=3$ for a tower scale $m_0^{(i)}\simeq \Lambda_{\text{c}}^{(i)}$ and UV cut-off $\Lambda_{\text{UV}}=\Lambda_{\text{QG}}$. The superconformal flavour models are attractive but the explicit realisations involve large strongly interacting groups and large number of extra chiral multiplets. This leads generically to the presence of Landau poles below the fundamental case. In our case that structure is replaced by towers of KK-like particles, which are common in string vacua, and induce large wave-function renormalisation for matter fields.
	
	\section{Flavour structure and neutrino masses}
	\label{s:EmergenceSM}
	
	If the emergence idea is correct, all kinetic terms of the particles in the SM should arise from loop corrections upon integration over infinite towers of states. However, the details of the resulting IR kinetic terms would strongly depend on which specific towers are coupled to the SM particles. There could be KK-like towers, string towers or both, whilst the SM massless particles could interact with different combinations of these towers. Thus, the kinetic terms will depend on the structure of the underlying theory and which particular point in moduli space the SM is sitting at. So one either has an specific compactification containing the SM in its massless sector and studies the different towers appearing as one moves in moduli space --- and how they couple to this massless sector --- or one makes some simplifying assumptions. We will follow here the latter approach with the aim of showing some possible phenomenological applications of Emergence to the SM physics as well as some interesting questions that arise in this attempt.

	\subsection{Emergence of hierarchical SM Yukawas}
	\label{ss:EmergenceYukawas}	
	
	Let us now consider possible effects on the structure of the Yukawas present in the Standard Model. We showed above how Emergence may naturally lead to hierarchies in Yukawa couplings through large wave-function renormalisation factors, $\mathcal{Z}_i$. We would now like to discuss whether this structure could provide in principle for an understanding of the hierarchies of fermion masses in the SM in terms of the structure of IR towers appearing in the underlying theory. In the emergence scheme all Yukawa couplings in the UV would be of order one, and the required hierarchies for the first two generations should arise because of large quantum corrections. This possibility has been considered in the past, thus e.g. in \cite{Georgi} it was proposed that `all fermions are created equal' and specific models were constructed in which the anomalous dimensions were large enough to cope with hierarchies of $\mathcal{O}(10^{-5})$, as required by the SM fermionic spectrum. However, this required from an extension of the SM which happened to be problematic, such that these attempts did not quite work in the end. Alternatively, one could try to obtain `hierarchical Yukawa couplings' upon starting with `hierachical metrics' in supergravity models, as considered e.g. in \cite{Ibanezcoset}. More recently, certain models were constructed which coupled the SM to a (nearly) SCFT sector in such a way that large anomalous dimensions were generated for the SM particles (see e.g. \cite{Nelson,Poland,Dudas}), as we mentioned above. We argue here that Emergence gives a new twist to these attempts: Large quantum corrections to the kinetic terms arise quite naturally because of the presence of relatively light towers interacting with the massless sector. Let us see how this could work in a $\mathcal{N}=1$ extension of the SM (although the main lessons would apply also to non-SUSY settings). In the UV, the relevant superpotential would have the form
	\beq
	W_Y\, =\,  {\overline H}\left(\mathcal{S}_{ij}^UQ_L^iU_R^j\, +\, \mathcal{S}_{ij}^{\nu}L^i\nu_R^j\right)\, +\, 
	H \left(\mathcal{S}_{ij}^D Q_L^iD_R^j\, +\, \mathcal{S}_{ij}^{E}L^i E_R^j \right)\, ,
	\eeq
	where the matrices $\mathcal{S}^{U,D,L,E}_{ij}$ have entries of order one and we use standard notation for the MSSM superfields (see e.g. \cite{Csaki:1996ks}). According to the above arguments, in the presence of towers generating the kinetic terms for all SM particles, one gets normalised Yukawa couplings
	\begin{equation}
		\begin{aligned}
			Y^U_{ij}\, &\simeq\,  \mathcal{S}_{ij}^U \left(g_{\bar H}g_Q^{(i)} g_U^{(j)}\right)^{-1/2}\, , \qquad Y^D_{ij} \simeq\,  \mathcal{S}_{ij}^D \left(g_{ H}g_Q^{(i)} g_D^{(j)}\right)^{-1/2}\, ,\\
			Y^E_{ij} &\simeq\,  \mathcal{S}_{ij}^E \left(g_{ H}g_L^{(i)} g_E^{(j)}\right)^{-1/2}\, , \qquad Y^\nu_{ij} \simeq \  \mathcal{S}_{ij}^\nu  \left(g_{\bar H}g_L^{(i)} g_{\nu_R}^{(j)}\right)^{-1/2}\, ,
		\end{aligned}
	\end{equation}
	where again we are taking the matter metric to be diagonal for simplicity. In order to recover the experimental fact that $m_{\text{top}}=m_U^{(3)} \simeq |\braket{\bar H}|\simeq 175$ GeV, we set $g_{\bar H} \simeq g_U^{(3)} \simeq g_Q^{(3)} \simeq 1$. For simplicity, we also impose $g_{H}\simeq g_D^{(3)} \simeq 1$, assuming that $m_{\text{bottom}}\ll m_{\text{top}}$ is due to the fact that $\tan\beta=|{\bar H}|/|H| \gg 1$, since this is enough to generate the desired hierarchies, but more involved models could also be considered. Define now small parameters
	\beq
	\label{eq:epsilondef}
	\epsilon_f^{{(i)}}\, =\, \left(g_f^{(i)}\right)^{-1/2}\, \qquad i=1,2\, , \qquad f=Q,U,D,L,\nu_R\, , 
	\eeq
	for the first two generations $i=1,2$. As a first approximation, let us consider the simple case in which these parameters only depend on the generation number so that $\epsilon_f^{{(i)}} = \delta_i$ for all $f$, such that we have only two of them, namely $\delta_1,\delta_2$. One can associate to those parameters certain infinite towers of states with characteristic mass scales $m_0^{(1)}, m_0^{(2)}$, such that one indetifies (c.f. eq. \eqref{eq:wavefunctionheavytowers})
	\beq
	\delta_1\, =\, \left(\frac {m_0^{(1)}}{\Lambda_{\text{QG}}} \right)^{3/2}\, , \qquad \delta_2\, =\, \left(\frac {m_0^{(2)}}{\Lambda_{\text{QG}}} \right)^{3/2}\, .
	\label{eq:deltas}
	\eeq
	Then we have a structure of fermion masses of the form
	\begin{equation}
		\begin{aligned}
			\notag	&\left(m_{\text{top}}, m_{\text{charm}}, m_{\text{up}}\right)  \sim |{\bar H}|\, \left(1,  (\delta_2)^2, (\delta_1)^2\right)\, , \qquad \left(m_\tau, m_\mu, m_e\right) \sim |{H}|\,\left(1,  (\delta_2)^2, (\delta_1)^2 \right)\, ,\\
			&\left(m_{\text{bottom}}, m_{\text{strange}}, m_{\text{down}}\right) \sim |{H}|\, \left(1,  (\delta_2)^2, (\delta_1)^2\right)\, , \qquad \left(m_{\nu_3}, m_{\nu_2}, m_{\nu_1}\right) \sim |{ \bar H}|\, \left(1,  (\delta_2)^2, (\delta_1)^2\right)\, ,
		\end{aligned}
	\end{equation}
	where we have set all the entries in the UV  couplings ${\cal S}_{ijk}$ to be of $\mathcal{O}(1)$. By choosing $\delta_1\sim 10^{-2}$ and $\delta_2\sim 10^{-1}$ one obtains qualitative agreement with the observed structure of fermion masses in the SM. Notice that eq. \eqref{eq:deltas} suggests us that the required towers would be at most one or two orders of magnitude below the QG scale $\Lambda_{\text{QG}}$. It can be checked that one also obtains an approximately diagonal CKM matrix, as observed experimentally. Of course, this is an oversimplified set-up and one may obtain more detailed agreement by playing e.g. with the order one parameters in the ${\cal S}$ entries or by considering e.g. leptons and quarks within the same generation to have different $\epsilon$-factors. The above considerations just illustrate how the notion of Emergence may provide us with new perspectives to address the flavour problem of the SM. In some way, it may be considered as a new `twist' to the old idea that flavour hierarchies may arise due to the presence of large flavour-dependent quantum corrections. In Emergence, there is a natural source for the required large wave-function renormalisation of quarks and leptons, namely the existence of infinite towers of states coupling to the light fermionic fields. The structure of masses observed at low energies would be a reflection of the structure of SM \emph{singlet} towers present at high energies. It would be interesting to explore in detail the new possibilities offered to fermion mass model-building from this renewed perspective. We refer the reader to \cite{Nelson,Poland} which even though they are based  on very different physics, lead to 
similar parameterisations.

	\begin{figure}[t]
		\begin{center}
			\subfigure[]{
				\includegraphics[height=3cm]{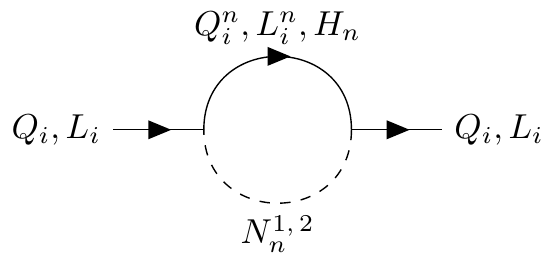}
			}\qquad 
			\subfigure[]{
				\includegraphics[height=3cm]{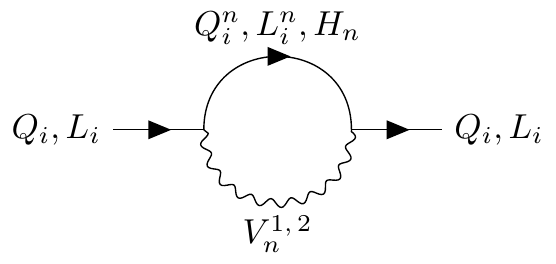}
			}\quad
			\subfigure[]{
				\includegraphics[height=3cm]{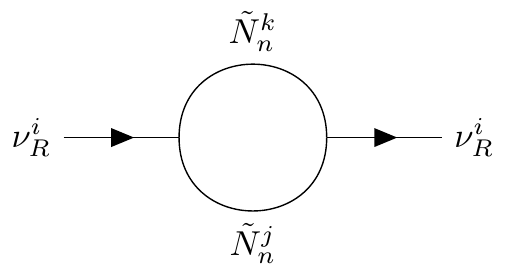}
				\label{kineticfermionsRHneutrinos}
			}
			\caption{One-loop diagrams  contributing to the kinetic terms of quarks and leptons. a) and b) from scalar or vector particles coupling to the first two generations ($i=1,\,2$).
				c) Right-handed neutrinos may get  additional large contributions from towers of SM singlets $N_n^i$.}
			\label{kineticfermions}
		\end{center}
	\end{figure}

	\subsubsection*{Dirac Neutrino masses and an extra dimension}	
	
	Neutrino masses are  very special in the context of the Standard Model, particularly due to their extremely small masses. We do not know yet what is the absolute value of the three neutrino masses, but oscillation data implies  for the  mass differences (see e.g. \cite{Gonzalez-Garcia:2021dve})
	\begin{equation}
	\Delta m_{21}^2\, =\, (7.42\pm 0.2)\times 10^{-5}\ \text{eV}^2\, , \qquad \Delta m_{31}^2\, =\, (2.51 \pm 0.028)\times 10^{-3}\ \text{eV}^2\, .
	\label{eq:deltamsquarednetrinos}
	\end{equation}
	With a normal hierarchy this suggests e.g. for the heaviest neutrino $m_{\nu_3}\simeq 5\times 10^{-2}$ eV. We do not know either whether they are of Dirac or Majorana type. Whereas the standard see-saw mechanism 
	\cite{Minkowski:1977sc,Gell-Mann:1979vob,Yanagida:1980xy} gives an elegant explanation to the smallness of neutrino Majorana masses, it is fair to say that an equally elegant mechanism for attaining small Dirac neutrino masses has not been proposed so far. In what follows, we describe how very small Dirac masses may appear in the context of Emergence, and in the next subsection we will show why those are required by the consistency of the UV theory, relating this to the smallness of the cosmological constant and the EW scale.
	
	We just saw that, in the presence of right-handed neutrinos, there will be Yukawa couplings given by
	\beq
	Y^\nu_{ij}\, \simeq\,  \mathcal{S}_{ij}^\nu  \left(g_{\bar H}g_L^{(i)} g_{\nu_R}^{(j)}\right)^{-1/2}\, .
	\eeq
	We set the metrics $g_{\bar H}\sim 1$ since, as discussed above, it qualitatively reproduces the known masses for SM fermions of the third generation. In order to obtain neutrino Dirac masses of order $10^{-1} - 10^{-3}$ eV one needs  Yukawa couplings of order $Y_\nu \sim 10^{-12}- 10^{-14}$, whereas $\left(g_L^{(i)}\right)^{-1/2}$ provides only for a small suppression factor if we want to recover the correct masses for the charged leptons. Thus, the only way to achieve such a large suppression is to have very large wave-function renomalisation effects for the right-handed neutrinos. Luckily, unlike the case of the rest of the SM fermions, the right-handed neutrinos can have couplings to (massive) singlets ${\tilde N}^i$ of the general form $\lambda_{ijk} (\nu_R^i {\tilde N}^j{\tilde N}^k)$ so that a diagram like that  in fig. \ref{kineticfermionsRHneutrinos}  will give rise to a large contribution to metric for the right-handed neutrinos if the towers of ${\tilde N}_i$ are light enough. Let us consider for simplicity the simplest case in which we have a single KK-tower of scale $m_0$ which couples equally to the three right-handed neutrinos. We also assume that right-handed neutrinos couple to the massive ${\tilde N}_n$ singlets with Yukawa couplings $Y_n \simeq n$, in analogy to how gauginos couple in a $\mathcal{N}=1$ SUSY theory. Then, from eq. \eqref{eq:U(1)gaugeEmergence} we will have neutrino Yukawa couplings
	\beq
	Y_\nu^{ij}\, \sim\,  {\cal S}_{ij}^\nu \frac {m_0}{M_p} \epsilon_L^{(i)}\, ,
	\label{eq:neutrinoYukawa}
	\eeq
	(no sum over indices), with $\epsilon_L^{(i)}$ defined in \eqref{eq:epsilondef}. Ignoring the order one factors one would get Dirac neutrino masses of the form
	\beq\label{eq:neutrinomasses}
	\left(m_{\nu_3},m_{\nu_2},m_{\nu_1}\right)\, \simeq\, M_{\text{EW}}\times \frac {m_0}{M_p}\, \left(1, \epsilon_L^{(2)}, \epsilon_L^{(1)} \right)\, ,
	\eeq
	where $M_{\text{EW}}=\left| \langle \bar H\rangle\right|$. Assuming normal neutrino hierarchy one estimates for the mass of the heaviest neutrino $m_{\nu_3}\simeq \sqrt{\Delta m^2_{32}}\simeq 5\times 10^{-2}$ eV. Thus imposing
	\beq
	m_{\nu_3}\, \simeq\, \frac {m_0}{M_p}M_{\text{EW}}\, \Rightarrow\, m_0\, \simeq\, 6.9\times 10^5\, \text{GeV}\, .
	\eeq
	So one concludes that in order to obtain Dirac neutrino masses at the observed range there must be a tower of states of characteristic scale $m_0\sim 7\times 10^5$ GeV. Above that energy, a 5th dimension opens up, being directly  felt only by right-handed neutrinos, since they are the only SM fields that couple to the tower.
	
	Let us remark that this mechanism to obtain very light Dirac neutrinos is slightly different to that considered in \cite{Dienes:1998sb}. In this reference the right-handed neutrinos are in the bulk of the theory, and get the small couplings due to the suppression of the volume of the bulk. In our case the right-handed neutrinos are not in the bulk of the theory and have no KK replicas. Rather they have couplings to singlets ${\tilde N}_n$ that are in the bulk of the theory and hence can have very light towers of KK states. The tower of states couples very weakly to the SM particles (only via the right-handed neutrinos, which couple themselves very weakly to the SM fields). With a mass of order $10^3$ TeV and such weak couplings it is not challenged by present 
	accelerator experiments, although it may be relevant in cosmology. Note in this connection that this kind of extra dimension is totally different to the large extra dimensions scenario of \cite{Arkani-Hamed:1998jmv} because the fundamental gravity scale is not just above the EW scale but much above. Indeed computing the species scale associated to this tower one obtains  $\Lambda_{\text{QG}} \sim  m_0^{1/3}M_p^{2/3}\simeq 10^{14}$ GeV.  Thus the fundamental scale in this theory is about two orders of magnitude below the standard SUSY-GUT scale $10^{16}$ GeV.

	\section{The hierarchy problem and the cosmological constant}
	\label{s:Hierarchy&cc}
	
	It has been argued in the context of the Swampland ideas that the mass of the lightest neutrino must be Dirac and that it is bounded from above by the cosmological constant scale, i.e.\cite{Ibanez:2017kvh,Ibanez:2017oqr,Hamada:2017yji,Gonzalo:2018tpb,Gonzalo:2018dxi,Gonzalo:2021zsp,Gonzalo:2021fma}
	\beq
	m_{\nu_1} \ \lesssim V_0^{1/4} \ .
	\label{eq:neutrinobound}
	\eeq
	As we said, 
	the origin of this constraint comes from the Non-SUSY AdS Conjecture of \cite{Ooguri:2016pdq}  and the AdS Distance Conjecture \cite{Lust:2019zwm}. The former states that there cannot be stable non-SUSY AdS vacua. It turns out that if one compactifies the SM on a circle \cite{Arkani-Hamed:2007ryu}, one finds that due to Casimir forces, an AdS minimum  appears in 3d unless the lightest neutrino has at least 4 degrees of freedom (i.e. it is Dirac) and mass $m_{\nu_1}\, \lesssim\, V_0^{1/4}$.\footnote{More precisely, circle compactification implies for neutrinos with normal hierarchy $m_{\nu_1}\leq 3.2\times V_0^{1/4}$ \cite{Ibanez:2017kvh}.} On the other hand, applying the latter to this 3d vacua also implies similar bounds \cite{Gonzalo:2021zsp}. Combining this expression with eq. \eqref{eq:neutrinomasses} yields the interesting result
	\beq
	\boxed {
		M_{\text{EW}}\,  m_0\, \epsilon_L^{(1)}\, \lesssim\,   V_0^{1/4}\, M_p }\, ,
	\label{geom}
	\eeq
	where here $m_0$ denotes the scale of the tower giving the leading large wave-function renormalisation to the lightest neutrino. Note that this implies an upper bound for the mass of the tower (at fixed EW scale)
	\beq
	m_0\, \lesssim\, \frac{V_0^{1/4}\, M_p }{\epsilon_L^{(1)}M_{\text{EW}} }\, \simeq\, \left(\epsilon_L^{(1)}\right)^{-1}10^4\, \text{GeV}\, .
	\eeq
	As we mentioned above, understanding the hierarchies of lepton masses suggests $\epsilon_L^{(1)}\simeq 10^{-2}$ so one gets the constraint $m_0\lesssim 10^6$ GeV, consistent with the previous estimation. This gives us an explanation of why Dirac masses are extremely small. They are so because otherwise the bound in eq. \eqref{eq:neutrinobound} would be violated, and in order to preclude this from happening, light towers of states ${\tilde N}$ must exist. This is very interesting because it gives an elegant explanation for the smallness of the neutrino masses. Furthermore, it also explains the experimental fact that the c.c. scale and  neutrino masses are very close to each other. The see-saw mechanism for Majorana neutrinos, on the contrary, does not provide for an explanation for this surprising experimental coincidence. Although this bound applies to the lightest neutrino mass, it is natural to expect the other two neutrino masses to be also tiny, given they will get a mass from an analogous mechanism, as discussed above.

	\begin{figure}[tb]
		\begin{center}
			\includegraphics[scale=0.27]{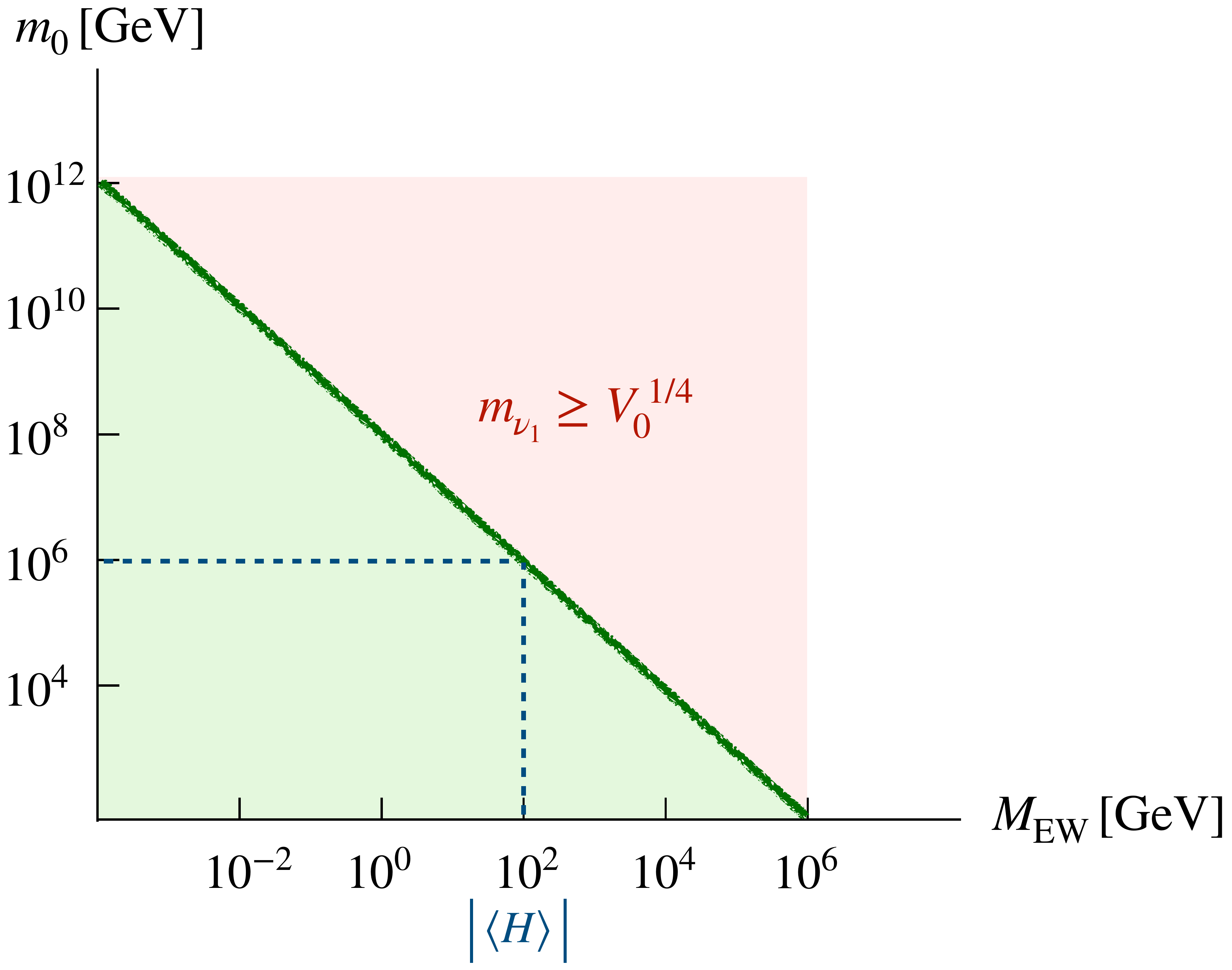}
			\caption{Limits on the plane of the tower scale $m_0$ versus the EW scale. The diagonal line corresponds to the bound in eq. \eqref{geom}. The neutrino mass spectrum fixes $m_0\simeq 7\times 10^5$ GeV, which in terms implies an upper bound on the EW scale of order $10^2$ GeV. On the other hand, a gravitino mass is bounded to be $m_{3/2}\lesssim 10^6$ GeV, implying a Higgs mass consistent with $m_{\text{H}}\simeq 126$ GeV. Note we have set $V_0^{1/4} \simeq 10^{-12}$ GeV in the plot, but the allowed region would actually shrink upon decreasing $V_0$.}
			\label{fig:plotm0vsMEW}
		\end{center}
	\end{figure}

	One can also obtain an upper bound for the mass  $m_0$ taking into account that the AdS non-SUSY constraint also tells us that the Higgs VEV cannot be smaller than $\Lambda_{\text{QCD}}\simeq 200$ MeV \cite{Gonzalo:2018dxi}, namely $M_{\text{EW}}\gtrsim \Lambda_{\text{QCD}}$.\footnote{The same constraint has been recently obtained from the Festina Lente bound \cite{Montero:2019ekk,Montero:2021otb}.} This comes about because,  when the Higgs contribution to quark masses is below the QCD scale, a number of pseudo-Goldstone scalars appear contributing negatively to the scalar potential, giving rise to unwanted AdS vacua en the SM compactified in the circle. Plugging this, together with $\epsilon_L^{(1)}\simeq 10^{-2}$, into eq. \eqref{geom} gives the following bound 
	\beq
	m_0\, \lesssim\, \frac{ V_0^{1/4}M_p }{\epsilon_L^{(1)}\Lambda_{\text{QCD}}}\, \simeq\, 10^{9}\, \text{GeV}\, .
	\label{QCD}
	\eeq
	Thus, once one accepts that SM Yukawa couplings to a SM  Higgs field exist, and the observed value of  $\Lambda_{\text{QCD}}$, this bound would force the theory to yield small neutrino Yukawa couplings, in order not to violate eq. \eqref{eq:neutrinobound}.
	
	Turning the argument around, given a tower with  mass scale fixed at  $m_0$, eq. \eqref{geom} implies an upper bound for the EW scale
	\beq
	M_{\text{EW}}\, \lesssim\, \left(\epsilon_L^{(1)}\right)^{-1} \times \frac {V_0^{1/4}M_p}{m_0}\,  \simeq\, 8 \times  \left(\epsilon_L^{(1)}\right)^{-1}\, \text{GeV}\, ,
	\label{higgsvev}
	\eeq 
	so that $M_{\text{EW}}\lesssim 8\times 10^2$ GeV (where we have used $m_0 \simeq 7 \times 10^5$ GeV). Thus the EW scale and the cosmological constant are thus correlated, and if $V_0\rightarrow 0$, also the EW scale will go down. This solves the Higgs hierarchy problem in the sense that any higher scale would lead to a violation of the  AdS conjectures above. These constraints on scales are summarised in fig. \ref{fig:plotm0vsMEW}. One observes that the EW and tower scales are close numerically, so from here on we assume that they scale in the same way with $V_0$, namely $m_0=\eta\, M_{\text{EW}}$ (with $\eta$ some $V_0$ independent parameter). We then obtain from eq. \eqref{geom} 
	\beq
		M_{\text{EW}}\, \lesssim\, \left(\epsilon_L^{(1)}\eta\right)^{-1/2} V_0^{1/8}M_p^{1/2}\, ,
	\eeq
	and also
	\beq
		m_0\, \lesssim\, \left(\left(\epsilon_L^{(1)}\right)^{-1}\eta\right)^{1/2}  V_0^{1/8}M_p^{1/2} \, .
	\eeq
	Interestingly, although we did not impose it a priori, the second  expression is in agreement with the AdS/dS conjecture of eq. \eqref{eq:ADC}, which states that as $V_0\rightarrow 0$, there must be a tower of states of scale $m_{\text{tower}}\lesssim V_0^\alpha M_p^{1-4\alpha}$, with $\alpha$ some positive number. In this case the conjecture is fulfilled with $\alpha=1/8$.\footnote{Eq. \eqref{QCD} naively seems to suggest that here $\alpha=1/4$. However, one should notice that the coefficient $M_p/\Lambda_{\text{QCD}}$ is indeed very large, and goes against the intuition that the coefficient in the AdS/dS conjecture should not be hierarchically large.} The fact that the bound in eq. \eqref{eq:neutrinoYukawa} implies a bound on the Higgs VEV was already pointed out in \cite{Ibanez:2017kvh,Ibanez:2017oqr}. Here however we associate the smallness of the Yukawa coupling to the existence of an extra dimension at a scale  of order $700$ TeV, leading to the explicit expression eq. \eqref{higgsvev}. A summary of the different scales in the theory (for a value of $\eta\simeq 10^4\, $, as required for our universe), are presented in table \ref{escalas}.

	\begin{table}[h!!]\begin{center}
			\renewcommand{\arraystretch}{1.00}
			\begin{tabular}{|c||c|c|c|}
				\hline
				Species\ scale &  $\Lambda_{\text{QG}}$  &  $\sim 10\times V_0^{1/24}M_p^{5/6}$  & $ \sim 10^{14}$\ GeV  \\
				\hline 
				Extra dimension& $m_0$  &    $\sim 10^{3} \times V_0^{1/8}M_p^{1/2}$  &   $\sim 10^{6}$\ GeV \\
				\hline
				Gravitino & $m_{3/2}$  &    $\lesssim  10^{3} \times V_0^{1/8}M_p^{1/2}$  &   $\lesssim 10^{6}$\ GeV \\
				\hline
				EW Scale &  $M_{\text{EW}} $ &  $\sim 10^{-1}\times V_0^{1/8}M_p^{1/2}$            &   $\sim 10^{2} $\ GeV\\
				\hline
				Dirac-$\nu$ & $m_{\nu_1}$  &  $\lesssim V_0^{1/4}$  &   $\lesssim 10^{-12}$\ GeV  \\
				\hline
				Cosmological constant & $V_0^{1/4}$ & $V_0^{1/4}$ &  $\simeq10^{-12}$\ GeV \\
				\hline
			\end{tabular}
			\caption{The value of the c.c., $V_0$, together with the Planck mass, dictate the size of  all fundamental scales.}
			\label{escalas}
		\end{center}
	\end{table}

	\subsubsection*{Some phenomenological implications}	
	
	It would be interesting to analyse possible particle physics and cosmological implications of the idea of Emergence and the existence of towers of singlet states around $m_0\sim 7\times 10^{5}$ GeV. Here we just point out some immediate phenomenological consequences of the above analysis.
	
	There are direct important implications for neutrino physics. In particular, in the present scheme neutrinos must be Dirac, in order to obey the $m_{\nu_1}\lesssim V_0^{1/4}$ Swampland constraint. Thus, neutrino-less double $\beta$-decays should yield negative results and, in addition, the lightest neutrino (in normal hierarchy) should verify ${m_{\nu_1}\leq 7.7\ \text{meV}}$ (see \cite{Ibanez:2017kvh,Ibanez:2017oqr,Gonzalo:2021zsp,Gonzalo:2021fma}). At the moment, the best results for the mass of the lightest neutrino come from the analysis of cosmological constraints combined with terrestrial data, see e.g. \cite{GAMBIT,olga,hannestad}. For instance, the results from \cite{olga} imply $m_{\nu_1}<20$ meV for normal hierarchy, which is only a factor 3 above the Swampland inspired bounds obtained in the aforementioned references. Thus, if evidence were found for a lightest neutrino mass around 20 meV, the Swampland bound $m_{\nu_1}\leq 7.7 $ meV would be challenged.	
	
	Another interesting question is baryon number violating processes, like proton decay in the present scheme. The species scale corresponding to this tower is given by $ \Lambda_{\text{QG}}\, \simeq\, m_0^{1/3}M_p^{2/3}\simeq 10^{14}$ GeV, a couple of orders of magnitude below the standard GUT scale $M_X\sim 10^{15}-10^{16}$ GeV. Then the rate for proton decay would apparently be larger than in ordinary GUT's by a factor $(M_X/\Lambda_{\text{QG}})^4\sim 10^8$, so that it should have been observed already. However this estimation ignores that the wave-function of the lightest quarks and leptons involved in dimension 6 proton decay operators, like $\left(Q_LQ_Lu_R^*(e^+_R)^*\right)$, further suppresses the (norm squared of the) amplitudes by powers of $\left(\epsilon_f^{(i)}\right)^8$. Therefore, the decay rates can easily be accommodated within the experimental limits. It would be interesting to make more precise determinations of the proton lifetime in specific models of fermion masses of the type discussed in section \ref{ss:EmergenceYukawas}. Baryon decay experiments can also potentially constrain the scheme discussed here.

	\subsubsection*{Supersymmetry and Mini-Split}	
	
	In all the above discussion supersymmetry does not play any crucial role, and it has been used at some points mainly to simplify the notation. Thus, the structure of scales arises as a consequence of Emergence and certain Swampland conjectures. In particular the EW scale stabilisation arises without supersymmetry. Still, there is the question of whether supersymmetry, if present at some scale in the infrared, would be restricted by similar arguments and what possible role it would play. In this regard, there are several low-energy relevant arguments in favour of a supersymmetric EFT. For instance, if the SM scalar potential is extrapolated to high energies, it is well known that it becomes unstable (or metastable) at a scale in the range of $10^9-10^{12}$ GeV \cite{Elias-Miro:2011sqh,Degrassi:2012ry}. This may be considered  as a potential problem for models that only consider the SM up to such scales. From the point of view of the Swampland, a scalar potential crossing the axis of zero energies may also imply the necessity of new physics in that region. A natural solution to this problem is to consider that at some scale below $10^9-10^{12}$ GeV, the SM becomes supersymmetric, in which case the potential is stabilised.

	Let us then assume that there is $\mathcal{N}=1$ SUSY below the fundamental scale $\Lambda_{\text{QG}}$. One can estimate the maximum value of the SUSY-breaking masses by considering the maximum possible value for the gravitino mass in the EFT below $\Lambda_{\text{QG}}$. We are going to assume that the superpotential in a hidden sector is bounded  as  $|W|_{\text{max}} \simeq \Lambda_{\text{QG}}^3$. A possible motivation for this is that specific models of moduli stabilisation in string theory have in general as an ingredient gaugino condensation, in which $|W| \simeq |\braket{\lambda\lambda}|\simeq \mu^3$, where $\mu$ is the condensation scale which must obviously verify $\mu \lesssim \Lambda_{\text{QG}}$. Then one can estimate\footnote{Note that, since in this case the gravitino mass would be related to the tower scale like $m_{3/2}\simeq m_0$, the Gravitino Conjecture of \cite{Cribiori:2021gbf, Castellano:2021yye} is fulfilled. This conjecture states that if in a class of theories $m_{3/2}\rightarrow 0$, a tower of particles should become massless with $m_0\lesssim m_{3/2}^\delta$. In the present case one has $\delta=1$.}
	\beq
	m_{3/2}\, \sim\, \frac {|W|}{M_p^2}\, \lesssim\, \frac {\Lambda_{\text{QG}}^3}{M_p^2}\, \simeq\, m_0\, \simeq\,  10^3\ V_0^{1/8}M_p^{1/2}\, ,
	\label{eq:gravitinomass}
	\eeq
	and thus $m_{3/2}  \simeq 6.9\times 10^2\ \text{TeV}$. Therefore, one expects a relatively low scale for the gravitino mass, and also for the SUSY spectrum. Note that a crucial ingredient to obtain this relatively low scale is that the fundamental scale has been lowered down to $\Lambda_{\text{QG}}\simeq 10^{14}$ GeV. Hence low-energy SUSY is related to the existence of the tower of scale $m_0$ coupled to neutrinos, which causes the reduction of the fundamental scale. Note also  that an additional a posteriori reason to assume that $|W|_{\text{max}} \simeq \Lambda_{\text{QG}}^3$ is that otherwise the particles in the tower coupled to neutrinos would get generically SUSY-breaking masses $M_{\text{SS}}>m_0$ and the computation of the large wave-function renormalisation of right-handed neutrinos could be spoiled.
	
	One would expect a SUSY spectrum of squarks and gluinos below that scale in the range $M_{\text{SS}}\simeq 1-10^3$ TeV. Interestingly it has been argued that the range of SUSY masses in that region is preferred by the measured value of the Higgs mass, $m_{\text{H}} \simeq 126$ GeV. The tree level value of the lightest Higgs in the MSSM is bounded by the $Z^0$ mass, $m_{Z^0}\simeq$ 91 GeV. It is well known that mass is increased by loop corrections of the qualitative form \cite{Ellis:1990nz,Haber:1990aw,Okada:1990vk}
	\beq
	\delta(m_{\text{H}}^2)\ \simeq \ \frac {3g_{\text{EW}}^2m_t^4}{4\pi^2M_W^2} \log\left(\frac {m_{\tilde t}}{m_t}\right) \ ,
	\eeq
	where $m_t,m_{\tilde t}$ are the top and stop masses, respectively. Thus getting a Higgs mass as high as $126$ GeV requires a relatively heavy squark SUSY spectrum. Specifically it has been argued that this value of the Higgs mass prefers a region of SUSY masses in the range $10^4-10^6$ GeV\cite{Arvanitaki:2012ps,Arkani-Hamed:2012fhg,Hall:2012zp}. In this 
	{\it Mini-Split}  scenario, the SUSY fermions (gauginos and Higgsinos) are lighter than the squarks and sleptons by a loop factor, so that they may be accessible at the LHC and/or the FCC. The lightest neutralinos would then provide natural candidates for dark matter. Concerning proton stability, dimension 6 operators are suppressed, as we already mentioned. Dimension 5 proton decay  operators are also suppressed both due to the relatively heavy SUSY spectrum   and the additional $\epsilon_f^{(1,2)}$  suppression factors that  both the fermions and sfermions carry. We direct the reader to \cite{Arvanitaki:2012ps,Arkani-Hamed:2012fhg,Hall:2012zp} for further phenomenological consequences of this Mini-Split scenario.

	The expression \eqref{eq:gravitinomass} above with the power $V_0^{1/8}$ is interesting and  has been proposed in the past (see \cite{Banks}), based on apparently different arguments.\footnote{See also \cite{Anchordoqui:2023oqm}, where the same power of the c.c. appears in the scaling of the SUSY breaking scale. Note, however, that this reference assumes the \emph{dark dimension} scenario, in which the scaling of the mass of the tower with the c.c. is $m_{0} \sim V_0^{1/4}$, as opposed to our $m_{0} \sim V_0^{1/8}$, yielding different values for exponents that fulfill the AdS Distance Conjecture and the Gravitino Conjecture in each scenario.} In this work it is argued that SUSY should be unbroken in flat space and that its breaking arises only due to the existence of a non-vanishing c.c., $V_0$. Then it is argued that $M_{\text{SUSY}}\sim\, V_0^\alpha$ and that the required $\alpha$ to agree with the observed EW scale (assuming $M_{\text{EW}}\sim m_{3/2}$) is $\alpha=1/8$. In our case such value for $\alpha$ arises because of the existence of a tower of states and a UV cut-off $\Lambda_{\text{QG}}\sim V_0^{1/24}$.

	Note that in the present scheme the EW scale is just a few orders of magnitude away from the SUSY-breaking scale, although the origin of both scales is not directly related. It is not SUSY which stabilises the Higgs scale, but the Quantum Gravity constraint that implies $m_{\nu_1}\lesssim V_0^{1/4}$. From the point of view of the EFT observer  the SUSY-breaking soft terms should appear (slightly)  fine-tuned in order to obtain the \emph{little hierarchy} $M_{\text{EW}}\ll m_{3/2}$. Note, however, that some gauginos and neutralinos  could have masses in the region $1-10$ TeV and might be accessible at the LHC and/or the FCC, as suggested in the Mini-Split scenario. As we said, this would also provide for adequate dark matter candidates. Concerning gauge coupling unification, since the fundamental scale $\Lambda_{\text{QG}}$ is two orders of magnitude below the standard SUSY-GUT scale of $10^{16}$ GeV, couplings would not unify sharply. On the other hand, one also expects the presence of threshold corrections from the flavour towers, which could improve this unification. Let us remark, nonetheless, that string theory vacua with SM spectrum  does not necessarily require  the presence of an $SU(5)$ symmetry at the fundamental scale. Notice, however, that the notion of unification of couplings suggested by the Emergence Proposal implies that all perturbative couplings in the IR should become large above $\Lambda_{\text{QG}}$, and thus they are in some sense unified \cite{Heidenreich:2017sim}.

	Before concluding, let us point out the main differences between the conclusions in this paper and the recently proposed \emph{dark dimension} scenario \cite{Montero:2022prj,Gonzalo:2022jac,Blumenhagen:2022zzw,Anchordoqui:2022tgp}.
	The results in the present article are based on essentially two principles: 1) Emergence of kinetic terms, which gives rise to hierarchies of  fermion masses (including small Dirac masses for neutrinos), and in turn implies the presence of a tower of singlet states at $m_0\simeq  7\times 10^5$ GeV; and 2) the AdS non-SUSY conjecture and/or the AdS Distance conjecture as applied to the 3d SM, which gives rise to constraints on the EW scale. On the other hand, in the \emph{dark dimension} scenario originally proposed in \cite{Montero:2022prj}, the starting point is to assume a tower at a scale ${m_{\text{tower}}\simeq V_0^{1/4}\sim 10^{-12} \ \text{GeV}}$, suggested by the AdS Distance Conjecture with the exponent $\alpha=1/4$. The neutrinos get a mass by mixing with the singlets in the towers and there can be deviations from Newton's law at short distances which are constrained by experiments.

	\section{The Upside Down Universe and entropy}
	\label{s:ccUniverse}
	\begin{figure}[tb]
		\begin{center}
			\includegraphics[scale=0.28]{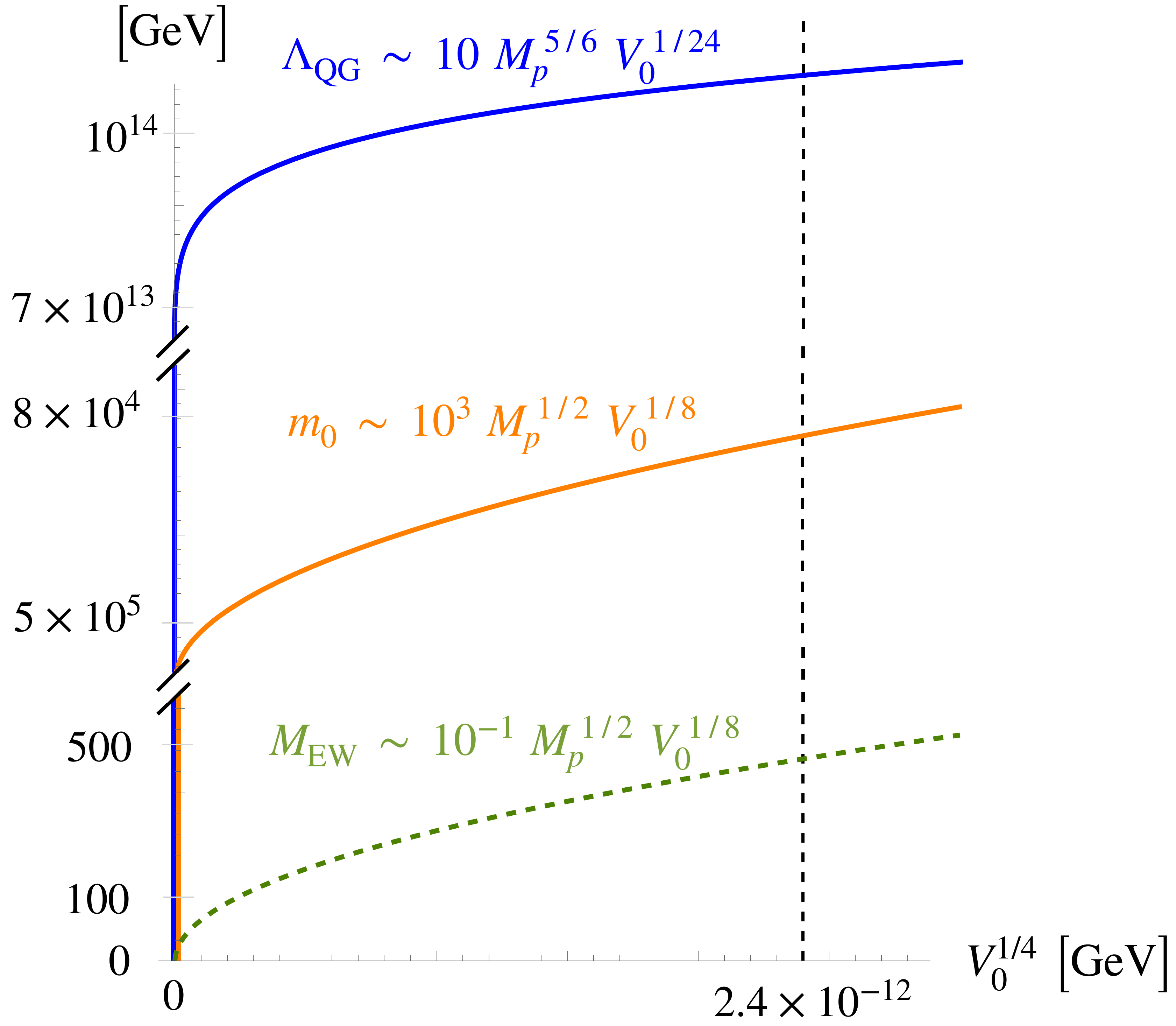}
			\caption{In the {\it Upside Down universe} all scales are dictated by the c.c., $V_0^{1/4}$. If the c.c. vanished all scales would become massless leading to infinite entropy. For a finite value of $V_0$, scales increase and get separated depending on different powers of $V_0$.}
			\label{fig:upside}
		\end{center}
	\end{figure}
	As summarized in table \ref{escalas}, the c.c. $V_0$ appears as the {\it fundamental scale}  in terms of which the other higher physical scales may be obtained, explicitly showing some sort of IR-UV connection. In this {\it Upside Down Universe}, when $V_0\rightarrow 0$, all fundamental scales in physics go to zero (in Planck units), see fig. \ref{fig:upside}. Note that we are not claiming that there cannot be masses or hierarchies in the presence of an exactly Minkowski vacuum ($V_0=0$), but our reasoning applies to families of vacua with non-vanishing $V_0$ in the limit $V_0 \to 0$. This suggests a change in how we address scales in fundamental physics. We are used to begin with the fundamental UV scale and then try to derive from it the relevant IR parameters, such as $M_{\text{EW}}$ and the c.c., ending up with an apparently extreme fine-tuning in order to accommodate the experimental results. On the other hand, given that all fundamental scales may be expressed in terms of $V_0$, perhaps one should take this as fundamental and the questions on fine-tunings are now replaced by \emph{scale separation} questions. Instead of asking why $M_{\text{EW}}\ll M_p$ we instead address the question why $M_{\text{EW}}\gg V_0^{1/4}$.  This resonates with some ideas in \cite{Banks} (see also \cite{Witten}) which imply that the c.c. is an input to the theory, instead of a parameter to be computed in field theory from the fields in the spectrum. The c.c. is in fact determined by the number of degrees of freedom in our de Sitter universe, and the dimension of the Hilbert space of the theory is  huge but nevertheless finite. The limit $V_0\rightarrow 0$ is a critical limit in which the number of degrees of freedom goes to infinity. It would be great to make the ideas from \cite{Banks} more precise, but in the meantime our conclusions seem to point towards similar directions. 
	
	\section{Outlook}
	\label{s:Outlook}
	
	In this paper  we have studied how the idea of \emph{Emergence} in Quantum Gravity may have implications for Standard Model physics. In its strongest formulation, this proposal posits that all kinetic terms are absent in the UV and are only generated in the IR via loop corrections involving towers of states becoming light (with respect to the UV scale). We have shown that indeed, all particles in the Standard Model may potentially get their kinetic terms from loop contributions associated to towers of massive states. We have also seen that hierarchies of Yukawa couplings may appear due to large quantum contributions to their kinetic terms induced by towers of \emph{singlets} coupling to the Standard Model fields. This is somewhat reminiscent of previous ideas which attempted to create such hierarchies by producing large anomalous dimensions for the quark and lepton fields. In Emergence the large anomalous dimensions are generated by one-loop contributions from towers of singlet states. 
	
	In the context of Emergence, Dirac neutrino masses are tiny because the  right-handed neutrinos may get particularly large contributions to their wave-function renormalisation from towers which have a low mass scale $m_{0}\simeq Y_{\nu_3}M_p \simeq 7\times 10^5$ GeV, with $Y_{\nu_3}\sim m_0/M_p\sim 10^{-12}$. The need to have such relatively light towers is forced upon us if we impose the AdS non-SUSY and/or the AdS/dS Distance Conjectures, which require $m_{\nu_1}\lesssim V_0^{1/4}$. This, unlike the case of the see-saw mechanism for Majorana masses, provides also for an explanation of the remarkable experimental coincidence between the (fourth root of the) cosmological constant and the neutrino mass scales. For the aforementioned value of the tower scale, $m_0\simeq 7\times 10^5\ \text{GeV} \, \left( \sim 10^3 \times V_0^{1/8}\, M_p^{1/2} \right)$,  the Electro-Weak scale is bounded from above as $M_{\text{EW}}\lesssim 10^2 \ \text{GeV} \, \left( \sim 10^{-1} \times V_0^{1/8}\, M_p^{1/2}\right)$. This connects the Electro-Weak hierarchy problem with the smallness of the cosmological constant in an interesting and novel way, and nicely connects with the idea that all the fundamental energy scales that we observe might be related to the cosmological constant, in such a way that they would vanish if the cosmological constant were set to zero \cite{Banks,Witten}.
	
	In the above scheme, there is an extra dimension appearing at an energy scale of order $1000$ TeV, and the right-handed neutrinos are the only Standard Model particles feeling it directly. Therefore, there are no measurable fifth forces. The tower which is becoming light as $V_0\to 0$ (according with the AdS/dS Distance Conjecture) is slightly above the Electro-Weak scale. Interestingly, the SUSY-breaking scale is also below $10^3$ TeV, which is a scale favoured by the  {\it Mini-Split} scenario of the  supersymmetric Standard Model, in which SUSY fermions are about two orders of magnitude below the bosonic masses. These would correspond to gauginos and neutralinos in the TeV range, accessible at the FCC and possibly at LHC. This is quite different as compared to the recently considered \emph{dark dimension} idea, in which the tower is of the order of the neutrino mass scale and fifth-forces may indeed arise (see \cite{Montero:2022prj, Anchordoqui:2022txe, Blumenhagen:2022zzw, Gonzalo:2022jac, Anchordoqui:2022tgp,Anchordoqui:2023oqm}). 
	
	The ideas discussed in this paper show possible ways in which Emergence may have a significant impact in Standard Model physics. The idea that in the UV the theory is strongly coupled or, equivalently, that there are no kinetic metrics to start with, implies that all non-trivial kinematics and scales in physics should appear as IR effects. For hierarchies to appear, either for charged fermions or neutrinos, strong wave-function renormalisation effects are needed, especially for the lightest particles. The case of Dirac neutrino masses is the most dramatic, requiring the existence of towers of singlet particles slightly above the Electro-Weak scale. It is quite non-trivial that phenomena including hierarchies of quarks and leptons, scales of neutrino masses, Electro-Weak scale (in terms of the cosmological constant), consistency with proton stability, understanding of the  value for the Higgs mass and low-energy SUSY all fit within the same scheme. Further study is clearly needed to elucidate phenomenological consequences of Emergence, as well as the interplay between that and the Swampland conditions.
	
	\vspace{1.5cm}
	\centerline{\bf \large Acknowledgments}
	\noindent We would like to thank E. Gonzalo, M. González-López D. L\"ust, F. Marchesano, M. Montero, A. Uranga and I. Valenzuela for useful discussions and correspondence. This work is supported through the grants CEX2020-001007-S, PGC2018-095976-B-C21 and PID2019-108892RB-I00, funded by MCIN/AEI/10. 13039/501100011033 and by ERDF A way of making Europe. The work of A.C. is supported by the Spanish FPI grant No. PRE2019-089790 and by the Spanish Science and Innovation Ministry through a grant for postgraduate students in the Residencia de Estudiantes del CSIC. The work of A.H. is supported by the ERC Consolidator Grant 772408-Stringlandscape.

	\bibliography{refs-emergencePHENO.bib}

\providecommand{\href}[2]{#2}\begingroup\raggedright\begin{thebibliography}{10}

\bibitem{Minkowski:1977sc}
P.~Minkowski, {\it {$\mu \to e\gamma$ at a Rate of One Out of $10^{9}$ Muon
  Decays?}},  {\em Phys. Lett. B} {\bf 67} (1977) 421--428.

\bibitem{Gell-Mann:1979vob}
M.~Gell-Mann, P.~Ramond, and R.~Slansky, {\it {Complex Spinors and Unified
  Theories}},  {\em Conf. Proc. C} {\bf 790927} (1979) 315--321,
  [\href{http://arxiv.org/abs/1306.4669}{{\tt arXiv:1306.4669}}].

\bibitem{Yanagida:1980xy}
T.~Yanagida, {\it {Horizontal Symmetry and Masses of Neutrinos}},  {\em Prog.
  Theor. Phys.} {\bf 64} (1980) 1103.

\bibitem{Elias-Miro:2011sqh}
J.~Elias-Miro, J.~R. Espinosa, G.~F. Giudice, G.~Isidori, A.~Riotto, and
  A.~Strumia, {\it {Higgs mass implications on the stability of the electroweak
  vacuum}},  {\em Phys. Lett. B} {\bf 709} (2012) 222--228,
  [\href{http://arxiv.org/abs/1112.3022}{{\tt arXiv:1112.3022}}].

\bibitem{Degrassi:2012ry}
G.~Degrassi, S.~Di~Vita, J.~Elias-Miro, J.~R. Espinosa, G.~F. Giudice,
  G.~Isidori, and A.~Strumia, {\it {Higgs mass and vacuum stability in the
  Standard Model at NNLO}},  {\em JHEP} {\bf 08} (2012) 098,
  [\href{http://arxiv.org/abs/1205.6497}{{\tt arXiv:1205.6497}}].

\bibitem{Feruglio:2015jfa}
F.~Feruglio, {\it {Pieces of the Flavour Puzzle}},  {\em Eur. Phys. J. C} {\bf
  75} (2015), no.~8 373, [\href{http://arxiv.org/abs/1503.04071}{{\tt
  arXiv:1503.04071}}].

\bibitem{Vafa:2005ui}
C.~Vafa, {\it {The String landscape and the swampland}},
  \href{http://arxiv.org/abs/hep-th/0509212}{{\tt hep-th/0509212}}.

\bibitem{Palti:2019pca}
E.~Palti, {\it {The Swampland: Introduction and Review}},  {\em Fortsch. Phys.}
  {\bf 67} (2019), no.~6 1900037, [\href{http://arxiv.org/abs/1903.06239}{{\tt
  arXiv:1903.06239}}].

\bibitem{vanBeest:2021lhn}
M.~van Beest, J.~Calder\'on-Infante, D.~Mirfendereski, and I.~Valenzuela, {\it
  {Lectures on the Swampland Program in String Compactifications}},  {\em Phys.
  Rept.} {\bf 989} (2022) 1--50, [\href{http://arxiv.org/abs/2102.01111}{{\tt
  arXiv:2102.01111}}].

\bibitem{Grana:2021zvf}
M.~Gra\~na and A.~Herr\'aez, {\it {The Swampland Conjectures: A Bridge from
  Quantum Gravity to Particle Physics}},  {\em Universe} {\bf 7} (2021), no.~8
  273, [\href{http://arxiv.org/abs/2107.00087}{{\tt arXiv:2107.00087}}].

\bibitem{Harlow:2015lma}
D.~Harlow, {\it {Wormholes, Emergent Gauge Fields, and the Weak Gravity
  Conjecture}},  {\em JHEP} {\bf 01} (2016) 122,
  [\href{http://arxiv.org/abs/1510.07911}{{\tt arXiv:1510.07911}}].

\bibitem{Grimm:2018ohb}
T.~W. Grimm, E.~Palti, and I.~Valenzuela, {\it {Infinite Distances in Field
  Space and Massless Towers of States}},  {\em JHEP} {\bf 08} (2018) 143,
  [\href{http://arxiv.org/abs/1802.08264}{{\tt arXiv:1802.08264}}].

\bibitem{Corvilain:2018lgw}
P.~Corvilain, T.~W. Grimm, and I.~Valenzuela, {\it {The Swampland Distance
  Conjecture for K\"ahler moduli}},  {\em JHEP} {\bf 08} (2019) 075,
  [\href{http://arxiv.org/abs/1812.07548}{{\tt arXiv:1812.07548}}].

\bibitem{Heidenreich:2017sim}
B.~Heidenreich, M.~Reece, and T.~Rudelius, {\it {The Weak Gravity Conjecture
  and Emergence from an Ultraviolet Cutoff}},  {\em Eur. Phys. J. C} {\bf 78}
  (2018), no.~4 337, [\href{http://arxiv.org/abs/1712.01868}{{\tt
  arXiv:1712.01868}}].

\bibitem{Heidenreich:2018kpg}
B.~Heidenreich, M.~Reece, and T.~Rudelius, {\it {Emergence of Weak Coupling at
  Large Distance in Quantum Gravity}},  {\em Phys. Rev. Lett.} {\bf 121}
  (2018), no.~5 051601, [\href{http://arxiv.org/abs/1802.08698}{{\tt
  arXiv:1802.08698}}].

\bibitem{Blumenhagen:2019qcg}
R.~Blumenhagen, D.~Kl\"awer, and L.~Schlechter, {\it {Swampland Variations on a
  Theme by KKLT}},  {\em JHEP} {\bf 05} (2019) 152,
  [\href{http://arxiv.org/abs/1902.07724}{{\tt arXiv:1902.07724}}].

\bibitem{Blumenhagen:2019vgj}
R.~Blumenhagen, M.~Brinkmann, and A.~Makridou, {\it {Quantum Log-Corrections to
  Swampland Conjectures}},  {\em JHEP} {\bf 02} (2020) 064,
  [\href{http://arxiv.org/abs/1910.10185}{{\tt arXiv:1910.10185}}].

\bibitem{EnriquezRojo:2020hzi}
M.~Enr\'\i{}quez~Rojo and E.~Plauschinn, {\it {Swampland conjectures for type
  IIB orientifolds with closed-string U(1)s}},  {\em JHEP} {\bf 07} (2020) 026,
  [\href{http://arxiv.org/abs/2002.04050}{{\tt arXiv:2002.04050}}].

\bibitem{Blumenhagen:2020dea}
R.~Blumenhagen, M.~Brinkmann, D.~Klaewer, A.~Makridou, and L.~Schlechter, {\it
  {KKLT and the Swampland Conjectures}},  {\em PoS} {\bf CORFU2019} (2020) 158,
  [\href{http://arxiv.org/abs/2004.09285}{{\tt arXiv:2004.09285}}].

\bibitem{Castellano:2022bvr}
A.~Castellano, A.~Herr\'aez, and L.~E. Ib\'a\~nez, {\it {The Emergence Proposal
  in Quantum Gravity and the Species Scale}},
  \href{http://arxiv.org/abs/2212.03908}{{\tt arXiv:2212.03908}}.

\bibitem{Marchesano:2022axe}
F.~Marchesano and L.~Melotti, {\it {EFT strings and emergence}},
  \href{http://arxiv.org/abs/2211.01409}{{\tt arXiv:2211.01409}}.

\bibitem{Georgi}
H.~Georgi, A.~E. Nelson, and A.~Manohar, {\it {On the Proposition That All
  Fermions Are Created Equal}},  {\em Phys. Lett. B} {\bf 126} (1983) 169--174.

\bibitem{Ibanezcoset}
L.~E. Ib\'a\~nez, {\it {Supersymmetric Coset Unified Theories}},  {\em Phys.
  Lett. B} {\bf 150} (1985) 127--132.

\bibitem{Nelson}
A.~E. Nelson and M.~J. Strassler, {\it {Suppressing flavor anarchy}},  {\em
  JHEP} {\bf 09} (2000) 030, [\href{http://arxiv.org/abs/hep-ph/0006251}{{\tt
  hep-ph/0006251}}].

\bibitem{Poland}
D.~Poland and D.~Simmons-Duffin, {\it {Superconformal Flavor Simplified}},
  {\em JHEP} {\bf 05} (2010) 079, [\href{http://arxiv.org/abs/0910.4585}{{\tt
  arXiv:0910.4585}}].

\bibitem{Dudas}
E.~Dudas, G.~von Gersdorff, J.~Parmentier, and S.~Pokorski, {\it {Flavour in
  supersymmetry: Horizontal symmetries or wave function renormalisation}},
  {\em JHEP} {\bf 12} (2010) 015, [\href{http://arxiv.org/abs/1007.5208}{{\tt
  arXiv:1007.5208}}].

\bibitem{Dvali:2009ks}
G.~Dvali and D.~Lust, {\it {Evaporation of Microscopic Black Holes in String
  Theory and the Bound on Species}},  {\em Fortsch. Phys.} {\bf 58} (2010)
  505--527, [\href{http://arxiv.org/abs/0912.3167}{{\tt arXiv:0912.3167}}].

\bibitem{Dvali:2007hz}
G.~Dvali, {\it {Black Holes and Large N Species Solution to the Hierarchy
  Problem}},  {\em Fortsch. Phys.} {\bf 58} (2010) 528--536,
  [\href{http://arxiv.org/abs/0706.2050}{{\tt arXiv:0706.2050}}].

\bibitem{Dvali:2010vm}
G.~Dvali and C.~Gomez, {\it {Species and Strings}},
  \href{http://arxiv.org/abs/1004.3744}{{\tt arXiv:1004.3744}}.

\bibitem{Ooguri:2016pdq}
H.~Ooguri and C.~Vafa, {\it {Non-supersymmetric AdS and the Swampland}},  {\em
  Adv. Theor. Math. Phys.} {\bf 21} (2017) 1787--1801,
  [\href{http://arxiv.org/abs/1610.01533}{{\tt arXiv:1610.01533}}].

\bibitem{Lust:2019zwm}
D.~L\"ust, E.~Palti, and C.~Vafa, {\it {AdS and the Swampland}},  {\em Phys.
  Lett. B} {\bf 797} (2019) 134867,
  [\href{http://arxiv.org/abs/1906.05225}{{\tt arXiv:1906.05225}}].

\bibitem{Arkani-Hamed:2007ryu}
N.~Arkani-Hamed, S.~Dubovsky, A.~Nicolis, and G.~Villadoro, {\it {Quantum
  Horizons of the Standard Model Landscape}},  {\em JHEP} {\bf 06} (2007) 078,
  [\href{http://arxiv.org/abs/hep-th/0703067}{{\tt hep-th/0703067}}].

\bibitem{Ibanez:2017kvh}
L.~E. Ib\'a\~nez, V.~Martin-Lozano, and I.~Valenzuela, {\it {Constraining
  Neutrino Masses, the Cosmological Constant and BSM Physics from the Weak
  Gravity Conjecture}},  {\em JHEP} {\bf 11} (2017) 066,
  [\href{http://arxiv.org/abs/1706.05392}{{\tt arXiv:1706.05392}}].

\bibitem{Ibanez:2017oqr}
L.~E. Ib\'a\~nez, V.~Martin-Lozano, and I.~Valenzuela, {\it {Constraining the
  EW Hierarchy from the Weak Gravity Conjecture}},
  \href{http://arxiv.org/abs/1707.05811}{{\tt arXiv:1707.05811}}.

\bibitem{Hamada:2017yji}
Y.~Hamada and G.~Shiu, {\it {Weak Gravity Conjecture, Multiple Point Principle
  and the Standard Model Landscape}},  {\em JHEP} {\bf 11} (2017) 043,
  [\href{http://arxiv.org/abs/1707.06326}{{\tt arXiv:1707.06326}}].

\bibitem{Gonzalo:2018tpb}
E.~Gonzalo, A.~Herr\'aez, and L.~E. Ib\'a\~nez, {\it {AdS-phobia, the WGC, the
  Standard Model and Supersymmetry}},  {\em JHEP} {\bf 06} (2018) 051,
  [\href{http://arxiv.org/abs/1803.08455}{{\tt arXiv:1803.08455}}].

\bibitem{Gonzalo:2018dxi}
E.~Gonzalo and L.~E. Ib\'a\~nez, {\it {The Fundamental Need for a SM Higgs and
  the Weak Gravity Conjecture}},  {\em Phys. Lett. B} {\bf 786} (2018)
  272--277, [\href{http://arxiv.org/abs/1806.09647}{{\tt arXiv:1806.09647}}].

\bibitem{Gonzalo:2021zsp}
E.~Gonzalo, L.~E. Ib\'a\~nez, and I.~Valenzuela, {\it {Swampland constraints on
  neutrino masses}},  {\em JHEP} {\bf 02} (2022) 088,
  [\href{http://arxiv.org/abs/2109.10961}{{\tt arXiv:2109.10961}}].

\bibitem{Arvanitaki:2012ps}
A.~Arvanitaki, N.~Craig, S.~Dimopoulos, and G.~Villadoro, {\it {Mini-Split}},
  {\em JHEP} {\bf 02} (2013) 126, [\href{http://arxiv.org/abs/1210.0555}{{\tt
  arXiv:1210.0555}}].

\bibitem{Arkani-Hamed:2012fhg}
N.~Arkani-Hamed, A.~Gupta, D.~E. Kaplan, N.~Weiner, and T.~Zorawski, {\it
  {Simply Unnatural Supersymmetry}},
  \href{http://arxiv.org/abs/1212.6971}{{\tt arXiv:1212.6971}}.

\bibitem{Hall:2012zp}
L.~J. Hall, Y.~Nomura, and S.~Shirai, {\it {Spread Supersymmetry with Wino LSP:
  Gluino and Dark Matter Signals}},  {\em JHEP} {\bf 01} (2013) 036,
  [\href{http://arxiv.org/abs/1210.2395}{{\tt arXiv:1210.2395}}].

\bibitem{PASCOS2022}
L.~E. Ib\'a\~nez, {\it {The Quantum Gravity Swampland and Particle Physics}},
  {\em Talk at PASCOS-2022, Heidelberg} (2022).

\bibitem{Banks}
T.~Banks, {\it {Cosmological breaking of supersymmetry?}},  {\em Int. J. Mod.
  Phys. A} {\bf 16} (2001) 910--921,
  [\href{http://arxiv.org/abs/hep-th/0007146}{{\tt hep-th/0007146}}].

\bibitem{Arkani-Hamed:2006emk}
N.~Arkani-Hamed, L.~Motl, A.~Nicolis, and C.~Vafa, {\it {The String landscape,
  black holes and gravity as the weakest force}},  {\em JHEP} {\bf 06} (2007)
  060, [\href{http://arxiv.org/abs/hep-th/0601001}{{\tt hep-th/0601001}}].

\bibitem{Harlow:2022gzl}
D.~Harlow, B.~Heidenreich, M.~Reece, and T.~Rudelius, {\it {The Weak Gravity
  Conjecture: A Review}},  \href{http://arxiv.org/abs/2201.08380}{{\tt
  arXiv:2201.08380}}.

\bibitem{Ooguri:2006in}
H.~Ooguri and C.~Vafa, {\it {On the Geometry of the String Landscape and the
  Swampland}},  {\em Nucl. Phys. B} {\bf 766} (2007) 21--33,
  [\href{http://arxiv.org/abs/hep-th/0605264}{{\tt hep-th/0605264}}].

\bibitem{Ooguri:2018wrx}
H.~Ooguri, E.~Palti, G.~Shiu, and C.~Vafa, {\it {Distance and de Sitter
  Conjectures on the Swampland}},  {\em Phys. Lett. B} {\bf 788} (2019)
  180--184, [\href{http://arxiv.org/abs/1810.05506}{{\tt arXiv:1810.05506}}].

\bibitem{Font:2019cxq}
A.~Font, A.~Herr\'aez, and L.~E. Ib\'a\~nez, {\it {The Swampland Distance
  Conjecture and Towers of Tensionless Branes}},  {\em JHEP} {\bf 08} (2019)
  044, [\href{http://arxiv.org/abs/1904.05379}{{\tt arXiv:1904.05379}}].

\bibitem{Obied:2018sgi}
G.~Obied, H.~Ooguri, L.~Spodyneiko, and C.~Vafa, {\it {De Sitter Space and the
  Swampland}},  \href{http://arxiv.org/abs/1806.08362}{{\tt arXiv:1806.08362}}.

\bibitem{Garg:2018zdg}
S.~K. Garg, C.~Krishnan, and M.~Zaid~Zaz, {\it {Bounds on Slow Roll at the
  Boundary of the Landscape}},  {\em JHEP} {\bf 03} (2019) 029,
  [\href{http://arxiv.org/abs/1810.09406}{{\tt arXiv:1810.09406}}].

\bibitem{Agrawal:2020xek}
P.~Agrawal, S.~Gukov, G.~Obied, and C.~Vafa, {\it {Topological Gravity as the
  Early Phase of Our Universe}},  \href{http://arxiv.org/abs/2009.10077}{{\tt
  arXiv:2009.10077}}.

\bibitem{Palti:2020tsy}
E.~Palti, {\it {Fermions and the Swampland}},  {\em Phys. Lett. B} {\bf 808}
  (2020) 135617, [\href{http://arxiv.org/abs/2005.08538}{{\tt
  arXiv:2005.08538}}].

\bibitem{Csaki:1996ks}
C.~Csaki, {\it {The Minimal supersymmetric standard model (MSSM)}},  {\em Mod.
  Phys. Lett. A} {\bf 11} (1996) 599,
  [\href{http://arxiv.org/abs/hep-ph/9606414}{{\tt hep-ph/9606414}}].

\bibitem{Gonzalez-Garcia:2021dve}
M.~C. Gonzalez-Garcia, M.~Maltoni, and T.~Schwetz, {\it {NuFIT: Three-Flavour
  Global Analyses of Neutrino Oscillation Experiments}},  {\em Universe} {\bf
  7} (2021), no.~12 459, [\href{http://arxiv.org/abs/2111.03086}{{\tt
  arXiv:2111.03086}}].

\bibitem{Dienes:1998sb}
K.~R. Dienes, E.~Dudas, and T.~Gherghetta, {\it {Neutrino oscillations without
  neutrino masses or heavy mass scales: A Higher dimensional seesaw
  mechanism}},  {\em Nucl. Phys. B} {\bf 557} (1999) 25,
  [\href{http://arxiv.org/abs/hep-ph/9811428}{{\tt hep-ph/9811428}}].

\bibitem{Arkani-Hamed:1998jmv}
N.~Arkani-Hamed, S.~Dimopoulos, and G.~R. Dvali, {\it {The Hierarchy problem
  and new dimensions at a millimeter}},  {\em Phys. Lett. B} {\bf 429} (1998)
  263--272, [\href{http://arxiv.org/abs/hep-ph/9803315}{{\tt hep-ph/9803315}}].

\bibitem{Gonzalo:2021fma}
E.~Gonzalo, L.~E. Ib\'a\~nez, and I.~Valenzuela, {\it {AdS swampland
  conjectures and light fermions}},  {\em Phys. Lett. B} {\bf 822} (2021)
  136691, [\href{http://arxiv.org/abs/2104.06415}{{\tt arXiv:2104.06415}}].

\bibitem{Montero:2019ekk}
M.~Montero, T.~Van~Riet, and G.~Venken, {\it {Festina Lente: EFT Constraints
  from Charged Black Hole Evaporation in de Sitter}},  {\em JHEP} {\bf 01}
  (2020) 039, [\href{http://arxiv.org/abs/1910.01648}{{\tt arXiv:1910.01648}}].

\bibitem{Montero:2021otb}
M.~Montero, C.~Vafa, T.~Van~Riet, and G.~Venken, {\it {The FL bound and its
  phenomenological implications}},  {\em JHEP} {\bf 10} (2021) 009,
  [\href{http://arxiv.org/abs/2106.07650}{{\tt arXiv:2106.07650}}].

\bibitem{GAMBIT}
{\bf Gambit Cosmology Workgroup} Collaboration, P.~St\"ocker et~al., {\it
  {Strengthening the bound on the mass of the lightest neutrino with
  terrestrial and cosmological experiments}},  {\em Phys. Rev. D} {\bf 103}
  (2021), no.~12 123508, [\href{http://arxiv.org/abs/2009.03287}{{\tt
  arXiv:2009.03287}}].

\bibitem{olga}
E.~Di~Valentino, S.~Gariazzo, and O.~Mena, {\it {On the most constraining
  cosmological neutrino mass bounds}},
  \href{http://arxiv.org/abs/2106.15267}{{\tt arXiv:2106.15267}}.

\bibitem{hannestad}
S.~Roy~Choudhury and S.~Hannestad, {\it {Updated results on neutrino mass and
  mass hierarchy from cosmology with Planck 2018 likelihoods}},  {\em JCAP}
  {\bf 07} (2020) 037, [\href{http://arxiv.org/abs/1907.12598}{{\tt
  arXiv:1907.12598}}].

\bibitem{Cribiori:2021gbf}
N.~Cribiori, D.~Lust, and M.~Scalisi, {\it {The gravitino and the swampland}},
  {\em JHEP} {\bf 06} (2021) 071, [\href{http://arxiv.org/abs/2104.08288}{{\tt
  arXiv:2104.08288}}].

\bibitem{Castellano:2021yye}
A.~Castellano, A.~Font, A.~Herraez, and L.~E. Ib\'a\~nez, {\it {A gravitino
  distance conjecture}},  {\em JHEP} {\bf 08} (2021) 092,
  [\href{http://arxiv.org/abs/2104.10181}{{\tt arXiv:2104.10181}}].

\bibitem{Ellis:1990nz}
J.~R. Ellis, G.~Ridolfi, and F.~Zwirner, {\it {Radiative corrections to the
  masses of supersymmetric Higgs bosons}},  {\em Phys. Lett. B} {\bf 257}
  (1991) 83--91.

\bibitem{Haber:1990aw}
H.~E. Haber and R.~Hempfling, {\it {Can the mass of the lightest Higgs boson of
  the minimal supersymmetric model be larger than m(Z)?}},  {\em Phys. Rev.
  Lett.} {\bf 66} (1991) 1815--1818.

\bibitem{Okada:1990vk}
Y.~Okada, M.~Yamaguchi, and T.~Yanagida, {\it {Upper bound of the lightest
  Higgs boson mass in the minimal supersymmetric standard model}},  {\em Prog.
  Theor. Phys.} {\bf 85} (1991) 1--6.

\bibitem{Anchordoqui:2023oqm}
L.~A. Anchordoqui, I.~Antoniadis, N.~Cribiori, D.~Lust, and M.~Scalisi, {\it
  {The Scale of Supersymmetry Breaking and the Dark Dimension}},
  \href{http://arxiv.org/abs/2301.07719}{{\tt arXiv:2301.07719}}.

\bibitem{Montero:2022prj}
M.~Montero, C.~Vafa, and I.~Valenzuela, {\it {The Dark Dimension and the
  Swampland}},  \href{http://arxiv.org/abs/2205.12293}{{\tt arXiv:2205.12293}}.

\bibitem{Gonzalo:2022jac}
E.~Gonzalo, M.~Montero, G.~Obied, and C.~Vafa, {\it {Dark Dimension Gravitons
  as Dark Matter}},  \href{http://arxiv.org/abs/2209.09249}{{\tt
  arXiv:2209.09249}}.

\bibitem{Blumenhagen:2022zzw}
R.~Blumenhagen, M.~Brinkmann, and A.~Makridou, {\it {The Dark Dimension in a
  Warped Throat}},  \href{http://arxiv.org/abs/2208.01057}{{\tt
  arXiv:2208.01057}}.

\bibitem{Anchordoqui:2022tgp}
L.~Anchordoqui, I.~Antoniadis, and D.~Lust, {\it {The Dark Universe: Primordial
  Black Hole $\leftrightharpoons$ Dark Graviton Gas Connection}},
  \href{http://arxiv.org/abs/2210.02475}{{\tt arXiv:2210.02475}}.

\bibitem{Witten}
E.~Witten, {\it {Quantum gravity in de Sitter space}},  in {\em {Strings 2001:
  International Conference}}, 6, 2001.
\newblock \href{http://arxiv.org/abs/hep-th/0106109}{{\tt hep-th/0106109}}.

\bibitem{Anchordoqui:2022txe}
L.~A. Anchordoqui, I.~Antoniadis, and D.~Lust, {\it {Dark dimension, the
  swampland, and the dark matter fraction composed of primordial black holes}},
   {\em Phys. Rev. D} {\bf 106} (2022), no.~8 086001,
  [\href{http://arxiv.org/abs/2206.07071}{{\tt arXiv:2206.07071}}].

\end{thebibliography}\endgroup
	\bibliographystyle{JHEP}

\end{document}